\documentclass[conference,compsoc]{IEEEtran}

\ifCLASSOPTIONcompsoc
  \usepackage[nocompress]{cite}
\else
  \usepackage{cite}
\fi
\usepackage{xurl} 
\usepackage{amsthm} 
\usepackage{amsmath,amssymb,amsfonts}
\usepackage{algorithm}
\usepackage{algpseudocode}
\usepackage{graphicx}
\usepackage{subcaption}
\usepackage{textcomp}
\usepackage{xcolor}
\usepackage{xspace}
\usepackage{url}
\usepackage{hyperref}
\usepackage{tocloft} 
\usepackage{float}   
\usepackage{framed}  
\usepackage{lipsum} 
\usepackage{multirow}
\usepackage{colortbl}
\usepackage{tikz-inet,pgf}
\usepackage{makecell}
\usepackage[framemethod=TikZ]{mdframed}
\usepackage{lineno}
\usepackage{siunitx}

\usepackage{tikz}

\usepackage{environ} 

\algrenewcommand\algorithmiccomment[1]{\hfill\textcolor{darkgray}{$\triangleright$~#1}}

\newif\ifshowproofs
\showproofstrue   
\NewEnviron{maybeProof}{
  \ifshowproofs
    \begin{proof}
      \BODY
    \end{proof}
  \fi
}

\NewEnviron{showWhenSubmit}{%
  \ifshowproofs
  \else
    \BODY
  \fi
}

\usetikzlibrary{matrix,calc,positioning}

\newtheorem{lemma}{Lemma}
\newtheorem{theorem}{Theorem}
\newtheorem{property}{Property}
\newtheorem{corollary}{Corollary}
\newtheorem{definition}{Definition}

\newtheorem{observation}{Observation}

\floatstyle{boxed}  
\newfloat{Functionality}{htbp}{lob}
\newfloat{Algorithm}{htbp}{loa}

\usepackage[normalem]{ulem}

\newcommand{\pubbr}{\textit{BR}}
\newcommand{\valbr}{\textit{BR}_V}

\newcommand{\indicator}{\mathcal{I}}
\newcommand{\tx}{\textit{tx}}
\newcommand{\txset}{\textit{TX}}
\newcommand{\pk}{\textit{PK}}
\newcommand{\sk}{\textit{SK}}

\newcommand{\report}{\textit{Rpt}}
\newcommand{\bid}{b}
\newcommand{\Includedtx}{B}
\newcommand{\rewarda}{R_{\textsf{publisher}}}
\newcommand{\rewardm}{R_{\textsf{validator}}}
\newcommand{\rewardr}{\textit{RV}}
\newcommand{\IncludeF}{\textbf{\textit{IncR}}}
\newcommand{\rewardb}{\textit{Bribe}}
\newcommand{\gentx}{\textit{PubR}}

\newcommand{\sortgentx}{\textit{SortPubR}}

\newcommand{\prover}{P}

\newcommand{\VRFGEN}{\textit{GEN}_{\textit{VRF}}}
\newcommand{\VRFVER}{\textit{VER}_{\textit{VRF}}}

\newcommand{\minr}{r_{\min}}
\newcommand{\maxr}{r_{\max}}
\newcommand{\vrev}{U_V}
 
\newcommand{\randomstring}{S}

\newcommand{\preinc}{\textbf{\textit{PrInc}}}
\newcommand{\pregen}{\textit{PubR}_{\textit{Prrr}}}

\newcommand{\argmax}{\mathop{\textit{argmax}}}

\newcommand{\maxtwo}[1]{\mathop{\max\nolimits^{(2)}}\limits_{#1}}

\newcommand{\IncR}{\textit{IncR}}
\newcommand{\TotalPubR}{\textit{RAllPub}}

\newcommand{\pubaction}[2]{\ensuremath{\gentx_{P#1}^{#2}}}

\newcommand{\brbaction}[2]{\ensuremath{\rewardb_{P#1}^{#2}}}

\newcommand{\pubsinglesteprev}[2]{\ensuremath{R_{P#1}^{#2}}}
\newcommand{\puballrev}[1]{\ensuremath{R_{P#1}}}
\newcommand{\pubutil}[1]{\ensuremath{U_{P#1}}}

\begin{document}

\author{\IEEEauthorblockN{Hongyin Chen}
\IEEEauthorblockA{Technion\\
hongyin.chen.contact@gmail.com}
\and
\IEEEauthorblockN{Yubin Ke}
\IEEEauthorblockA{Peking University\\
keyubin097@gmail.com}
\and
\IEEEauthorblockN{Xiaotie Deng}
\IEEEauthorblockA{CFCS, Peking University\\
xiaotie@pku.edu.cn}
\and
\IEEEauthorblockN{Ittay Eyal}
\IEEEauthorblockA{Technion\\
ittay@technion.ac.il}
}

\title{Prrr: Personal Random Rewards for Blockchain Reporting\textsuperscript{*}
}

\sloppy

\maketitle

\begingroup
\renewcommand{\thefootnote}{*}
\makeatletter
\renewcommand{\@makefntext}[1]{%
  \noindent\makebox[1.5em][l]{\@makefnmark}#1}
\makeatother
\footnotetext{Published in the 2026 IEEE Symposium on Security and Privacy (S\&P). Please cite as~\cite{anonymous2025prrr}.}
\endgroup

\begin{abstract}
Smart contracts, the stateful programs running on blockchains, often rely on \emph{reports}.
\emph{Publishers} are paid to submit these reports on the blockchain.
Designing protocols that incentivize timely reporting is the prevalent \emph{reporting problem}. 
But existing solutions face a security-performance trade-off: 
Relying on a small set of trusted publishers introduces centralization risks, while allowing open submission results in an excessive number of reports on the blockchain.
We identify the root cause of this trade-off to be the standard symmetric reward design, which treats all reports equally.
We prove that no symmetric-reward mechanism can overcome the trade-off.

We present \emph{Personal Random Rewards for Reporting} (\emph{Prrr}), a protocol that assigns random heterogeneous values to reports.
We call this novel mechanism-design concept \emph{Ex-Ante Synthetic Asymmetry}.
To the best of our knowledge, Prrr is the first game-theoretic mechanism (in any context) that deliberately forms participant asymmetry.
Prrr employs a second-price-style settlement to allocate rewards, ensuring incentive compatibility and achieving both security and efficiency.
Following the protocol constitutes a Subgame-Perfect Nash Equilibrium, robust against collusion and Sybil attacks.
Prrr is applicable to numerous smart contracts that rely on timely reports.
\end{abstract}

\IEEEpeerreviewmaketitle



\section{Introduction}

Blockchain \emph{smart contracts} are stateful programs that facilitate the \$$170$B Decentralized Finance ecosystem~\cite{coindesk2025defitvl}.
A wide variety of contracts require \emph{reports} with external data for their functionality~\cite{moosavi2023fast,motepalli2023sok,wang2025p,zamyatin2019xclaim,herlihy2018atomic,xie2022zkbridge,miller2019sprites,gudgeon2020sok,khabbazian2019outpost,liu2020fail,avarikioti2020cerberus,he2023don,qin2021empirical,morini2017managing,dai2025arbitrage,chen2024perpetual}.
In all cases, a smart contract requires a report to trigger its functionality. 
Principals called \emph{publishers} should publish reports and blockchain operators, called \emph{validators}, should include a report in the blockchain. 
But, as is the nature of decentralized systems, participants might not follow the protocol; they take the actions that maximize their profits. 
The smart contract therefore pays the publishers with cryptocurrency, and the challenge is to correctly incentivize the desired behavior. 
We call this the \emph{Blockchain Reporting Problem}.

%

Previous work~(\S\ref{sec:relatedwork}) and a variety of deployed systems face this problem. 
Many pay a fixed reward to the publisher who generated the first report placed on chain. 
Others include multiple reports on chain and distribute a reward among them. 
However, this approach creates a security-performance trade-off: Relying on a small set of publishers introduces a single point of failure, as demonstrated in recent incidents~\cite{cointelegraph2025base,coindesk2023arbitrum,cointelegraph2025starknet}. 
Conversely, allowing anyone to publish reports results in intense competition.
This erodes the participation incentive or leads to an excessive number of on-chain reports (even in protocols that reward just one report), as observed in liquidation events~\cite{daian2020flash,qin2021empirical,cointelegraph2021feeskyrocket}. 
On Ethereum~\cite{buterin2013ethereum}, there are over 1,000 daily reports, with their costs reaching approximately 14 times the median transaction cost in their respective blocks (\S\ref{appendix:ethereum_data}).

To analyze the reporting problem, we model~(\S\ref{sec:model}) a smart contract requesting one report.
There are two types of rational participants: \emph{publishers} and \emph{blockchain validators}. 
Validators generate blocks that contain \emph{transactions}. 
Publishers generate reports and send them to validators in transactions along with fees and bribes.
The set of publishers is unbounded and unknown to the contract in advance.
The validators choose which reports to include in the block. 

Our goal is to design a reporting protocol that ensures \emph{efficiency}, i.e., the number of included reports is upper bounded by a constant~$C$; \emph{decentralization}, i.e., participants have positive revenue, incentivizing their participation; \emph{fairness}, i.e., participants are rewarded in proportion to their efforts, to avoid centralization; and \emph{progress}, i.e., reports are included in a timely manner, preventing censorship.

We find that the standard symmetric mechanism design principle~(\S\ref{sec:impossibility}) is the root cause of the security-performance trade-off of previous solutions.
This commonly used principle treats all reports symmetrically, meaning any two reports are interchangeable.
We prove that, for any symmetric mechanism satisfying efficiency (for any~$C$), fairness, and progress, if the total number of generated reports exceeds $C$, all publishers earn zero revenue, violating decentralization.

Practical observations align with this impossibility result.
Consider the widely-used symmetric mechanism that allocates a fixed reward to the first report.
It creates a priority gas auction (PGA) among publishers, where they bribe validators, each competing to include their report first.
Empirical research demonstrates that when PGA bidders (publishers in our model) can bribe validators, they engage in a race-to-the-bottom, where most revenue flows to validators, and bidders' returns approach zero~\cite{mclaughlin2023large}.

We therefore propose \emph{Ex-Ante Synthetic Asymmetry} (\emph{EASA}, \S\ref{sec:mechanism}), a mechanism-design concept that induces ex-ante randomness to create controlled asymmetry. 
To the best of our knowledge, EASA is the first game-theoretic approach that deliberately forms participant asymmetry.

Based on EASA, we present \emph{Personal Random Rewards for Reporting} (\emph{Prrr}), a reporting protocol that assigns random heterogeneous values to reports with a specific random-value function.
Prrr requires publishers to publish all reports without offering bribes to the validators. 
To incentivize both participant types to follow the protocol, Prrr requires the validator to include in its block the reports with the highest and second-highest random values. 
The contract then rewards the validator with an amount equal to the value of the second-highest report, and it rewards the publisher who published the highest-value report with a reward equal to the difference between its value and the second-highest value.
Security thus relies on a careful choice of the random-value function. 
It should ensure that publisher rewards are monotonic in the number of published reports and should ensure validators include transactions, even in the face of bribes.
We provide two functions that satisfy these requirements.

As is typical in decentralized algorithms~\cite{tsabary2021mad,brugger2023checkmate,yaish2023uncle,mirkin2020bdos,carlsten2016instability}, the security of Prrr relies on its incentive compatibility. 
Prrr gives rise to a sequential game among the publishers and the validators~(\S\ref{sec:game}).
Each step of the game consists of two phases. 
In the first phase, publishers publish reports and offer bribes to the validator. 
In the second phase, the validator selects a vector of reports to include in the block based on the received reports and bribes.

We analyze the game~(\S\ref{sec:ic}) using backward induction and demonstrate that adhering to the Prrr protocol constitutes a Subgame-Perfect Nash Equilibrium.
We further show the robustness of this equilibrium: 
First, collusion among publishers and Sybil attacks are not beneficial. 
Second, collusion between publishers and the validator in the current step
does not yield higher utility for the colluding parties. 
Third, any deviation by validators strictly reduces their own utility (with one of our random-value functions). 
And finally, if a publisher's deviation impacts the revenue of other publishers, the deviating publisher's own revenue strictly decreases.

In all, our main contributions are: 
(1)~Modeling of the reporting problem; 
(2)~Proof that no symmetric mechanism satisfies our desired properties;
(3)~A novel mechanism-design concept that assigns heterogeneous values to reports via ex-ante randomness;
(4)~Prrr, a second-price-style reward allocation rule using the above concept;
(5)~Carefully chosen random-value functions for Prrr; and
(6)~Game-theoretic analysis showing that Prrr is a robust equilibrium.

Prrr is directly applicable to numerous smart contracts that rely on reports (see examples in Appendix~\ref{appendix:practical_conversions}). 
It enables, for the first time, decentralizing this central piece of the blockchain ecosystem.

\section{Related Work}\label{sec:relatedwork}

Although numerous works consider stochastic settings in mechanism design (see~\cite{borgers2015introduction} and, e.g.,~\cite{hartline2009simple,bergemann2010dynamic,myerson1981optimal}), to the best of our knowledge, we are the first to deliberately impose participant asymmetry to achieve desired incentives.



Many blockchain smart contracts rely on external data to maintain their functionality.
It arises in a variety of applications like Rollups~\cite{moosavi2023fast,donno2022optimistic,motepalli2023sok,wang2025p}, cross-chain protocols~\cite{zamyatin2019xclaim,herlihy2018atomic,ou2022overview,xie2022zkbridge}, misbehavior reporting protocols~\cite{miller2019sprites,poon2016bitcoin,
gudgeon2020sok,xu2023silentower,khabbazian2019outpost,liu2020fail,avarikioti2020cerberus,he2023don,he2022contract}, and financial events~\cite{qin2021empirical,morini2017managing,dai2025arbitrage,chen2024perpetual}.

All these settings share the same essential feature: a smart contract requires \emph{at least one participant} to publish a report in order to trigger a procedure.
The smart contract allocates rewards to publishers and publishers compete for rewards when publishing such reports.
Mechanisms without such competition fall outside our scope. 
For example, Proof-of-Work NFT~\cite{pownft2025} mints and other parallel minting schemes let each submitter independently create or claim value, so there is no race over the same report. 
Likewise, collaborative aggregation systems (e.g., Oracles~\cite{eskandari2021sok,breidenbach2021chainlink,adler2018astraea,zhang2020deco}) seek quorum or averaging rather than compete on publishing functionally-equivalent reports, and thus do not fit our model.

Most existing mechanisms for the blockchain reporting problem pay a fixed bounty to the first valid report. 
This winner-takes-all rule creates a security-performance trade-off.
If a protocol designates a small set or even a single publisher, authority concentrates and a single point of failure emerges. 
This raises manipulation and censorship risk and has caused real outages in rollups~\cite{cointelegraph2025base,coindesk2023arbitrum,cointelegraph2025starknet}.
Conversely, if the protocol allows any publisher to publish reports (e.g., liquidation protocols), publishers race to be first, leading to congestion and higher fees in practice~\cite{daian2020flash,qin2021empirical,cointelegraph2021feeskyrocket}.

This trade-off cannot be resolved by fair transaction ordering, which aims to sequence blockchain transactions based on the order of their publication~\cite{kursawe2020wendy,orda2021enforcing,kelkar2020order,kelkar2022order,cachin2022quick}. 
A publisher with superior connectivity or computational power can consistently publish first, securing rewards and discouraging others, ultimately driving the system toward centralization.

Some works pursue decentralized report \emph{generation} in rollups~\cite{motepalli2023sok,wang2025p}, but the reporting problem is about \emph{publication and inclusion} on chain, which these designs do not address.

Other efforts target reporting within particular domains.
ZkFair~\cite{zkfaird} rewards multiple publishers for publishing the same claim, but it forgoes efficiency, as every report must be recorded on chain. 
Chainlink~\cite{breidenbach2021chainlink} adopts a watchdog-priority scheme for price-bias alerts: Publishers receive sequential opportunities to publish an alert, and if a higher-priority node does not publish, the chance passes to the next.
Mouallem et al.~\cite{mouallem2026resilient} formalize this setting as the \emph{alerting problem} and design protocols whose reward and penalty rules maximize the budget a bribing adversary must spend to suppress an alert.
Both, however, assume a predefined set of staked publishers, whereas our model allows public participation.


        \section{The Reporting Problem}\label{sec:model}

We present the participants of the reporting problem~(\S\ref{sec:model:participants}), the blockchain model~(\S\ref{sec:model:blockchain}), how the system progresses~(\S\ref{sec:model:progress}), how reports are distributed~(\S\ref{sec:model:report}), and the goal of the reporting protocol~(\S\ref{sec:model:goal}).


            \subsection{Participants}\label{sec:model:participants}

The system comprises two sets of participants: a set ${\textbf{\prover} = \{\prover_1, \prover_2, \ldots, \prover_n\}}$ of 
\emph{Publishers} and a set $\textbf{V}$ of blockchain \emph{validators}.
Publisher~$\prover_j$ generates \emph{reports}, each with a fixed \emph{cost} $c_j$.
She can generate a limited number of reports within a time interval $t$, which we denote by~$N_j^{\max}(t)$.
Note that we do not predefine the set of publishers and allow for public participation.
The definition of sets is used solely for analytical purposes.
Decentralized systems can enforce both rate-limiting and uniqueness of reports using two main approaches: Proof-of-Work (PoW) and Proof-of-Stake (PoS). 
In PoW, generating reports incurs computational costs. 
The system enforces rate-limiting through computational power and ensures uniqueness by requiring a nonce for each report~\cite{aleo2025posvpow}. 
In PoS, generating reports requires financial stake.
The system enforces rate-limiting based on the amount of stake and ensures uniqueness by assigning a unique index to each report.

Each validator $V\in \textbf{V}$ holds a public key $\pk$ and a private key $\sk$.
The public key is common knowledge and only she knows her private key.
Publishers and validators are rational and aim to maximize their revenue.

            \subsection{Blockchain}\label{sec:model:blockchain}

The blockchain is a replicated state machine whose transitions are recorded as a ledger, a sequence of log entries called \emph{transactions}. 
\emph{Smart contracts} are stateful programs that run on the blockchain. 
They are triggered by transactions, execute arbitrary logic, and update their internal state. 
They can also receive and send virtual tokens~(cryptocurrency). 
Validators execute a state machine replication~(consensus) protocol, extending the list of blockchain transactions by appending batches of transactions called \emph{blocks} to the blockchain.
Validators take turns generating the blocks and appending them to the blockchain, forming a total order of blocks. 
We denote the block in position (index) $k$ by $\Includedtx^k$ and the validator who generates it by $V^k$.
We omit the details of the blockchain protocol and focus on the block contents.

Before $V^k$ generates block $\Includedtx^k$, an external source provides a random beacon pulse $Q^k$.
No one has any information about this beacon before the appending of block $\Includedtx^{k-1}$ or can influence its randomness.
And after the appending of block $\Includedtx^{k-1}$, the beacon pulse is public and becomes common knowledge.
Such randomness~\cite{das2022spurt,bhat2021randpiper} is already in use for blockchain mechanism design~\cite{chung2023foundations}.
To allow flexibility in the use of randomness, validator $V^k$ generates an additional unbiased pseudorandom string $S^k$ from the beacon pulse~$Q^k$.
We do not impose restrictions on how $S^k$ is generated; it may be identical to~$Q^k$ or derived with other methods chosen by the validator.
Hereafter, we refer to $S^k$ as the random string.

The $k$-th block is a tuple containing its index~$k$, the random beacon pulse $Q^k$, the random string $S^k$, and a vector of transactions ${\txset^k = (\tx_1, \tx_2, \cdots)}$, i.e., ${\Includedtx^k = (k, Q^k, S^k, \txset^k)}$.
A reporting transaction~$\tx$ is a pair of a data structure \emph{report}, $\report$, and a non-negative value $\bid$ representing the amount of fees the participant who generated the transaction is willing to pay for validators to include the transaction, i.e., $(\report, \bid)$.
For simplicity, we only consider reporting transactions in this paper, ignoring unrelated transactions. 
We refer to $\bid$ as the \emph{transaction fee bid} of transaction $\tx$.

\subsection{System Progress}\label{sec:model:progress}

A smart contract requires a report to maintain its functionality within a time window spanning $T$ blocks.
We call the time window an \emph{epoch}.
There can be either regularly repeated sequential epochs (e.g., Rollups~\cite{starknet2023, polygonzkEVMDocs}) or a single epoch (e.g., Watchtowers~\cite{khabbazian2019outpost}).
In practical systems, a smart contract might handle multiple epochs simultaneously for independent processes.
However, in this discussion, we focus on the scenario involving a single epoch, as repetitions are multiple independent instances of a single epoch.

Publishers put each report in a transaction and send it to validators.
Validators maintain a local set of transactions.
Upon receiving a transaction, a validator appends it to the set.
When generating a block  $\Includedtx^k$, validator~$V^k$ selects transactions from her local set of transactions as the transaction list~$\txset^k$ in the block $\Includedtx^k$.
If $\tx \in \txset^k$, we say that $V^k$ includes~$\tx$.

Publishers can also directly bribe validators. 
This bribery enables publishers to influence the contents of blocks to maximize their rewards. 
Formally, a publisher~$\prover_j$ provides a bribe function to validators, which we denote by~$\rewardb_{\prover j}$.
It takes a block $\Includedtx$ as input and outputs the bribe amount~$\rewardb_{\prover j}(\Includedtx)$ for the validator if she generates the block.
Practical systems like Flashbots~\cite{li2023demystifying} already facilitate such payments.

The system progresses in steps within an epoch, where each step corresponds to a block.
Within each step, first publishers generate and send transactions and offer bribes to validators, and then validators generate a block and append it to the blockchain.

        \subsection{Reporting Protocol}\label{sec:model:report}

The \emph{reporting protocol} indicates how publishers publish reports, validators include reports, and the smart contract distributes rewards to them.
Starting from a certain step within the epoch, participants execute the following protocol every step until the end of this epoch, which consists of three phases: \emph{Publication}, \emph{Inclusion}, and \emph{Processing}.

\subsubsection{Publication Phase}
Each publisher~$\prover_j$ has generated a set of reports $\textbf{\report}_j$.
In this phase, each publisher $\prover_j$ forms a set of transactions $\gentx_j$ from her reports $\textbf{\report}_j$ and generates a bribe function $\rewardb_{Pj}$. 
Then, publisher~$\prover_j$ publishes $(\gentx_j, \rewardb_{Pj})$ to validators.


\subsubsection{Inclusion Phase}

When a validator $V^k$ starts to generate a block, she executes the inclusion phase of the reporting protocol.
She first generates the random string $S^k$ from the beacon pulse~$Q^k$ and selects a set of transactions~$\txset^k$ from her local transaction set.
Then, she takes~$Q^k$,~$S^k$ and~$\txset^k$ to create the block $\Includedtx^k = (k, Q^k, S^k, \txset^k)$ and appends it to the blockchain.

The number of validators is large, so the validator~$V^k$ in each block~$k$ within the epoch is distinct and has distinct interests.

\subsubsection{Processing Phase}

When a validator $V^k$ appends a block~$\Includedtx^k$ to the blockchain, the smart contract executes the processing phase of the reporting protocol. 
This phase takes the vector of transactions $\txset^k$ of block $\Includedtx^k$ and the random string $S^k$ as input and outputs the payment for reports and the validator.

\subsection{Goal}\label{sec:model:goal}

A protocol solves the reporting problem if it satisfies \emph{decentralization}, \emph{fairness}, \emph{efficiency} and \emph{progress}.
We define these properties as follows.

We do not rely on a predefined set of publishers.
Therefore, decentralization means publishers are motivated to join the system. 
\begin{definition}[Decentralization]
A reporting protocol is \emph{decentralized} if all participants have positive expected revenue ignoring the cost of generating reports.
\end{definition}

Fairness ensures that publishers with many reports cannot gain additional rewards by splitting their identity into multiple smaller publishers. 
It also ensures that multiple publishers cannot increase their rewards by pooling together as a single publisher.

\begin{definition}[Fairness]
A reporting protocol is \emph{fair} if the expected revenue ignoring the cost of generating reports for each publisher is proportional to the number of reports she generates and is independent of how the total reports are distributed among publishers.
\end{definition}

Both decentralization and fairness are crucial to incentivize participants to join the system.

Efficiency indicates that there are not excessive on-chain reports which cause inefficiency in the underlying blockchain system:
\begin{definition}[$C$-Efficiency]
A reporting protocol is \emph{efficient} if validators include at most a constant number $C$ of reports in total in an epoch, regardless of how many reports publishers generate.
\end{definition}

The progress property ensures that the smart contract can process reports quickly, reflecting the censorship-resistance of the protocol.
\begin{definition}[Progress]
A reporting protocol satisfies \emph{progress} if the validator includes at least one report at the first step where participants start executing the reporting protocol.
\end{definition}

A smart contract executes the processing phase as prescribed. 
However, publishers and validators might deviate from the protocol to maximize their revenue.
Since we cannot force participants to follow the protocol, ensuring security requires that the protocol design makes deviations unprofitable. 
In other words, the protocol must be incentive-compatible, that is, following the protocol is an equilibrium.

        \section{Impossibility under Symmetric Design}\label{sec:impossibility}

In the reporting problem, all valid reports are functionally equivalent.
This naturally leads to the consideration of \emph{symmetric reporting protocols}, which treat all reports equally, independent of the publisher's identity. 
We show that no incentive-compatible symmetric reporting protocol can achieve all of efficiency, decentralization and progress.

\subsubsection*{Symmetric Reporting Protocol}

Before formally defining symmetry, we introduce some notation.

If a publisher publishes a bid for a transaction, it is equivalent to offering an additional bribe contingent on the transaction's inclusion.
Therefore, without loss of generality, we merge bids into bribes and consider only bribes in the following discussion.
Correspondingly, we abstract away transactions and only consider reports in the publication and inclusion.

We now formalize the reward and bribe functions.
Given a vector of reports $\textbf{\report} = (\report_1, \report_2, \dots, \report_l)$ and a random string $S$ as input, the publisher reward function outputs a vector of rewards for each report, ${\rewarda(\textbf{\report}, S) = (r_1, r_2, \dots, r_l)}$ and the validator reward function outputs a reward for the validator, $\rewardm(\textbf{\report}, S) = r_V$.
Given a vector of reports $\textbf{\report}$ and a random string $S$ as input, the bribe function of a publisher outputs a non-negative bribe amount $\rewardb(\textbf{\report}, S) $ for the validator.

A symmetric protocol should have symmetric reward for reports, which treats all reports equally.
This means that replacing any report in the input vector with another report should not change the output of the reward function.

\begin{definition}[Symmetric Reward]\label{def:symmetric_reward}
    For any two vectors of reports of equal length, $\textbf{\report}$ and $\textbf{\report}^{\prime}$, and for any random string $S$, the reward functions allocate rewards identically: $\rewarda(\textbf{\report},S) = \rewarda(\textbf{\report}^{\prime},S)$ and $\rewardm(\textbf{\report},S) = \rewardm(\textbf{\report}^{\prime},S)$.
\end{definition}

We aim to design a reporting protocol that excludes bribes (and bids). 
However, since we are proving an impossibility, we allow bribes in the protocol design to strengthen the result. 
We still impose symmetry constraints on such bribes since the protocol is symmetric.
Note that transaction fee bids included in bribes do not break the symmetry, as the bid for a report is paid only when the report is included.

\begin{definition}[Symmetric Bribery] \label{def:symmetric_bribery}
For any two vectors of reports of equal length, $\textbf{\report} = (\report_1, \dots, \report_l)$ and $\textbf{\report}^{\prime} = (\report_1^{\prime}, \dots, \report_l^{\prime})$, and for any random string $S$, the bribe function of publisher~$\prover_j$ is identical if her reports are in the same positions in both vectors, that is, if ${\forall 1 \leq i \leq l}$: ${\report_i \in \textbf{\report}_{Pj} \vee \report_i^{\prime} \in \textbf{\report}_{Pj}}$ implies $\report_i = \report_i^{\prime}$, then ${\rewardb_{Pj}(\textbf{\report}, S) = \rewardb_{Pj}(\textbf{\report}^{\prime}, S)}$.
\end{definition}

Note that publishers can still choose to deviate to an asymmetric bribery function since the protocol cannot enforce their actions.


\subsubsection*{Symmetric Game}
Assume there exists a symmetric reporting protocol satisfying efficiency, fairness, progress, and incentive compatibility. 
We aim to demonstrate that if any publisher has positive expected revenue, another publisher can deviate in a way that strictly increases their expected revenue.
This contradicts incentive compatibility. 
To establish this, we first formalize the game-theoretic model of a general symmetric reporting protocol.

Since the protocol satisfies progress, the validator includes at least one report in the first step.
There are two types of deviations.
First, a publisher may attempt to bribe the validator to exclude all reports in this step.
Second, a publisher may deviate to increase her own expected revenue while still ensuring that the validator includes at least one report in this step.
For our impossibility proof, we focus on constructing the second type of deviation.
After the first step, no participant can obtain any further reward from the smart contract, regardless of their actions.
Recall that we already assume that validators have independent interests~(\S\ref{sec:model:report}), the validator in the first step is only concerned with her revenue for that step.

Therefore, it suffices to construct the game in the first step with the second type of deviation.
Publishers first publish their transactions and bribe functions, after which the validator selects which transactions to include.
This interaction forms a multi-leader, single-follower Stackelberg game.

The actions of the participants are as follows.
For any publisher~$\prover_j$ with a set of reports ${\textbf{\report}_{Pj}}$, her action consists of selecting a subset of reports $\gentx_{Pj} \subseteq \textbf{\report}_{Pj}$ to publish, along with a bribe function ${\rewardb_{Pj}}$ offered to the validator.
The validator's action is to select an ordered vector ${\IncludeF}$ of reports to include in the block. 
The validator chooses this vector from the union of all published reports ${\bigcup_{1 \leq j \leq n} \gentx_{Pj}}$, and determines the final set and order of reports included in the block.

Denote by $R(\IncludeF, \textbf{\report}, S)$ the total reward allocated by the smart contract to the set of reports $\textbf{\report}$, given the included report vector ${\IncludeF = (\IncR_1, \IncR_2, \ldots )}$ and random string $S$.
Let $\rewarda(i,\textbf{\report}, S)$ be the reward assigned to the $i$-th report $\report_i$ in $\textbf{\report}$, that is, $i$-th element in the vector $\rewarda(\textbf{\report}, S)$.
Then, $R$ is the sum of the rewards for all reports in $\IncludeF$ that also belong to $\textbf{\report}$:
\begin{equation*}
R(\IncludeF, \textbf{\report}, S) = \sum_{\substack{\IncR_i \in \IncludeF \;\&\;  \\ \IncR_i \in \textbf{\report}}} \rewarda(i,\textbf{\report}, S) \,\,.
\end{equation*}

We now define the revenue for each participant.
Since the cost of generating reports for each publisher is fixed and independent of their actions within the game, we omit it from the revenue definition.
Given the random string~$S$, the action of a publisher~$\prover_j$, ($\gentx_{Pj}, \rewardb_{Pj}$), and the validator's action $\IncludeF$, the revenue of~$\prover_j$ is the total reward from her included reports minus her bribe paid to the validator:
\begin{multline}\label{eq:impossibility:pub_rev}
R_{Pj}(\gentx_{Pj},\rewardb_{Pj}, \IncludeF,   S) = \\ 
R(\IncludeF, \gentx_{Pj}, S) - \rewardb_{Pj}(\IncludeF, S) \,\,.
\end{multline}

The utility of a publisher~$\prover_j$ is her expected revenue over the randomness of $S$, denoted by $E_{S}[\cdot]$.

\begin{multline*}
U_{Pj}(\gentx_{Pj},\rewardb_{Pj}, \IncludeF) = \\ E_{S}[R_{Pj}(\gentx_{Pj},\rewardb_{Pj}, \IncludeF, S)] \,\,.
\end{multline*}

For the validator, given the vector of bribe functions of all publishers ${\textbf{\rewardb} = (\rewardb_{P1}, \ldots, \rewardb_{Pn})}$, her action~$\IncludeF$ and the random string $S$, her revenue is the reward from the smart contract plus the sum of bribes from all publishers:
\begin{multline}\label{eq:impossibility:val_rev}
R_{V}(\textbf{\rewardb}, \IncludeF, S) = \\
\rewardm(\IncludeF, S) + \sum_{1\leq j \leq n} \rewardb_{Pj}(\IncludeF, S)\,\,.
\end{multline}

\subsubsection*{Impossibility under Symmetry}

We show that no symmetric reporting protocol can achieve all of our desired goals.

\begin{theorem}\label{Theorem:impossibility}
For any reporting protocol with symmetric rewards and symmetric bribery, if it satisfies incentive compatibility, efficiency (i.e., includes at most $C$ reports), and progress, then whenever the total number of reports generated by all publishers exceeds $C$, all publishers have zero expected revenue.
\end{theorem}
Intuitively, suppose the total number of reports generated by all publishers is $N > C$.
By the fairness property, the expected revenue for each report is the same as in the case where $N$ publishers each generate one report.
Assume, for contradiction, that the expected revenue for each report is positive, so at least one publisher has positive revenue for some execution with a random string~$S$.
However, since~$N > C$, at least one publisher must have no report included by the validator in this execution.
We construct a deviation for this excluded publisher: She offers a small bribe to replace the report of a publisher with a positive revenue, capturing part of that positive revenue.
This profitable deviation contradicts incentive compatibility.
\begin{showWhenSubmit}The proof is in the full version of this report~\cite{anonymous2025prrr}.\end{showWhenSubmit}

\begin{maybeProof}
Assume there exists an instance of the game in which the total number of reports generated by all publishers is~$N > C$, and the expected revenue for at least one publisher is positive.
By the fairness property, the expected revenue for each report is also positive and identical to the case where~$N$ publishers each generate one report, since the total number of reports is the same.
Therefore, it suffices to show that if there are $N > C$ publishers, each generating one report, and the expected revenue for each report is positive, then the protocol cannot be incentive-compatible.

We prove by contradiction. 
It suffices to construct a deviation for any publisher and a corresponding best response for the validator such that the publisher's revenue increases and the validator's action is uniquely optimal.
In other words, if the publisher deviates to the new action, the validator's only best response results in strictly higher revenue for the publisher.
This contradicts the assumption that the protocol is incentive-compatible.

For any publisher~$\prover_j$ where $1 \leq j \leq n$, we denote her only report by $\report_{Pj}$.
She should publish it, i.e., ${\gentx_{Pj} = \{\report_{Pj}\}}$.
Otherwise, she would have zero revenue, contradicting the assumption that the expected revenue for each generated report is positive.

Without loss of generality, we assume publisher $\prover_1$ has a positive expected revenue.
It means that there exists a value of the random string, $S^*$, such that her revenue is positive, denoted by $R_1^* > 0$.

In this case, we denote the best response taken by the validator by ${\IncludeF^* = (\IncR^*_1, \IncR^*_2, \ldots)}$.
Due to efficiency, the validator includes at most $C$ reports.
Therefore, at least one publisher has no report included by the validator in this case.
Without loss of generality, we denote this publisher by~$\prover_2$.
So $\prover_2$ gets zero reward in this case:
\begin{equation}\label{eq:impossibility:pj_zero}
R(\IncludeF^*, \{\report_{P2}\}, S^*) = 0 \,\,.
\end{equation}

We first construct a deviation for publisher~$\prover_2$ that increases her revenue and find the validator's best response following this deviation.
Then, we calculate the revenue of publisher~$\prover_2$ showing that the deviation is profitable for her.

(i) \textbf{Deviation and Validator's Best Response}.

We construct a deviation letting $\prover_2$ seize $\prover_1$'s revenue~$R_1^*$ with the random string $S^*$.

Since $\IncludeF^*$ is the validator's best response with random string $S^*$, the validator has the maximal revenue for choosing $\IncludeF^*$:
\begin{equation}\label{eq:impossibility:original_br}
\forall \IncludeF:  R_{V}(\textbf{\rewardb}, \IncludeF, S^*) \leq R_{V}(\textbf{\rewardb}, \IncludeF^*, S^*) \,\,.
\end{equation}

Denote by $\IncludeF^{\prime}$ the report vector that is identical to the original validator's best response $\IncludeF^*$ but with~$\prover_1$'s report~$\report_{P1}$ replaced by~$\prover_2$'s previously unincluded report~$\report_{P2}$.

Publisher $\prover_2$ constructs a new bribe function $\brbaction{2}{\prime}$.
Given the random string $S^*$, the new bribe function is identical to her original bribe function $\rewardb_2$ for all possible included vectors, except for~$\IncludeF^{\prime}$.
Publisher $\prover_2$ pays the bribe to $\IncludeF^{\prime}$ which is the sum of the bribes that both $\prover_1$ and $\prover_2$ originally paid for $\IncludeF^*$, plus ($1/3$ of $R_1^*$):
\begin{multline}\label{eq:impossibility:dev_bribe}
    \brbaction{2}{\prime}(\IncludeF, S^*) =\\
    \begin{cases}
    \rewardb_{P2}(\IncludeF^*,S^*) +  &  \\
    \;\;\rewardb_{P1}(\IncludeF^*,S^*) + R_1^*/3 & \text{if } \IncludeF = \IncludeF^{\prime}; \\
    \rewardb_{P2}(\IncludeF,S^*) & \text{otherwise.}
    \end{cases}
\end{multline}

Now, we analyze the validator's revenue.
We denote by ${\textbf{\rewardb}^{\prime} = (\rewardb_{P1},\brbaction{2}{\prime}, \rewardb_{P3}, \ldots, \rewardb_{Pn})}$ the vector of all publishers' bribes of the deviation case.
The validator's revenue for choosing the report vector $\IncludeF$ with the new bribe functions and the random string $S^*$ is~(Equation~\ref{eq:impossibility:val_rev}):
\begin{align}\label{eq:impossibility:rev_new_act}
    &R_{V}(\textbf{\rewardb}^{\prime}, \IncludeF^{\prime}, S^*) \nonumber\\
    =& \rewardm(\IncludeF', S^*) + \rewardb_{P2}'(\IncludeF',S^*) + \nonumber\\
    &\sum_{j \neq 2} \rewardb_{Pj}(\IncludeF',S^*) \nonumber\\
    =& \rewardm(\IncludeF', S^*) + \rewardb_{P2}(\IncludeF^*, S^*) +R_1^*/3 +  \nonumber\\
    &  \rewardb_{P1}(\IncludeF^*, S^*)  +  \sum_{j \neq 2} \rewardb_{Pj}(\IncludeF',S^*)\,\,.
\end{align}

Now we consider the validator's revenue for choosing any report vector $\IncludeF$ other than $\IncludeF^{\prime}$.
Given the new bribe functions and the random string $S^*$, the validator's revenue for choosing the report vector $\IncludeF$ is (Equation~\ref{eq:impossibility:val_rev}):
\begin{align*}
    &R_{V}(\textbf{\rewardb}^{\prime}, \IncludeF \neq \IncludeF^{\prime}, S^*) \\
    =& \rewardm(\IncludeF, S^*) + \rewardb_{P2}'(\IncludeF,S^*) + \\ & \sum_{j \neq 2} \rewardb_{Pj}(\IncludeF,S^*) \\
    \stackrel{Eq. \ref{eq:impossibility:dev_bribe}}{=}& \rewardm(\IncludeF, S^*) + \rewardb_{P2}(\IncludeF,S^*) +  \\ 
    &\sum_{j \neq 2} \rewardb_{Pj}(\IncludeF,S^*) \\
    =& R_{V}(\textbf{\rewardb}, \IncludeF, S^*) + \sum_{1 \leq j \leq n} \rewardb_{Pj}(\IncludeF,S^*)
\end{align*}

This is exactly the revenue of the validator for choosing the original best response $\IncludeF$ to the original bribe functions $\textbf{\rewardb}$ and the random string $S^*$.
Combining with Equation~\ref{eq:impossibility:original_br}, we have that the revenue of the validator for choosing any report vector $\IncludeF$ other than $\IncludeF^{\prime}$ with the new bribe function is upper bounded by the revenue for choosing the original best response $\IncludeF^*$ with the original bribe functions:
\begin{equation}\label{eq:impossibility:val_rev_upperbound}
R_{V}(\textbf{\rewardb}^{\prime}, \IncludeF \neq \IncludeF^{\prime}, S^*) \leq R_{V}(\textbf{\rewardb}, \IncludeF^*, S^*) \,\,. 
\end{equation}

Combining the above inequality with Equations~\ref{eq:impossibility:val_rev} and~\ref{eq:impossibility:rev_new_act}, we calculate the difference in the validator's revenue between choosing $\IncludeF^{\prime}$ and any other report vector $\IncludeF$ with the deviation bribe functions $\textbf{\rewardb}^{\prime}$ and the random string~$S^*$:
\begin{align*}
& R_{V}(\textbf{\rewardb}^{\prime}, \IncludeF^{\prime}, S^*) -  \\
&R_{V}(\textbf{\rewardb}^{\prime}, \IncludeF \neq \IncludeF^{\prime}, S^*) \\
\stackrel{Eq. \ref{eq:impossibility:val_rev_upperbound}}{\geq}& R_{V}(\textbf{\rewardb}^{\prime}, \IncludeF^{\prime}, S^*) - R_{V}(\textbf{\rewardb}, \IncludeF^*, S^*) \\ 
\stackrel{Eq. \ref{eq:impossibility:val_rev}, \ref{eq:impossibility:rev_new_act}}{=}& (\rewardm(\IncludeF', S^*) -  \rewardm(\IncludeF^*, S^*))  + \\
& \rewardb_{P1}(\IncludeF^*, S^*) + R_1^*/3+\\
&\sum_{j \neq 2} (\rewardb_{Pj}(\IncludeF',S^*) - \rewardb_{Pj}(\IncludeF^*, S^*)) \,\,.
\end{align*}

Since the only difference between $\IncludeF'$ and~$\IncludeF^*$ is the inclusion of $\report_{P2}$ instead of $\report_{P1}$, due to the symmetric reward property, we have ${\rewardm(\IncludeF', S^*) = \rewardm(\IncludeF^*, S^*)}$.
Similarly, for any other publisher~${P_j \neq P_1, P_2}$, their reports in $\IncludeF^*$ and $\IncludeF'$ are identical.
Due to symmetric bribery, we have $\rewardb_{Pj}(\IncludeF',S^*) = \rewardb_{Pj}(\IncludeF^*,S^*)$.
Therefore, we can simplify the above difference to:
\begin{multline*}
R_{V}(\textbf{\rewardb}^{\prime}, \IncludeF^{\prime}, S^*) - R_{V}(\textbf{\rewardb}^{\prime}, \IncludeF \neq \IncludeF^{\prime}, S^*) =\\ \rewardb_{P1}(\IncludeF^*, S^*) + R_1^*/3.
\end{multline*}

Since bribes are non-negative and we assumed $R_1^* > 0$, the difference is strictly positive. 
The validator's revenue from any other choice remains unchanged. 
Thus, the validator's unique best response to the deviation bribe functions~$\textbf{\rewardb}^{\prime}$ and the random string $S^*$ is to select the report vector~$\IncludeF'$.

We have already shown that the deviation makes $\IncludeF^{\prime}$ the unique best response with random string $S^*$.
We should then show that this deviation strictly increases publisher~$\prover_2$'s expected revenue.
However, before that, we need to ensure that the deviation does not alter the validator's best response for any other random string $S' \neq S^*$.
With random string~$S'$, the validator may have multiple best responses with identical revenue.
The validator possibly selects different best responses related to $S'$ with the original bribe functions~$\textbf{\rewardb}$ and with deviating bribe functions~$\textbf{\rewardb}^{\prime}$, even though these two bribe functions are identical for the random string~$S'$.
This could affect publisher~$\prover_2$'s revenue with the deviation.

To resolve this, we can let $\prover_2$ slightly increase the bribe for the original best response with $S'$ so that it becomes uniquely optimal. As a result, for all random strings other than $S^*$, the validator's best response to the deviated bribe functions $\textbf{\rewardb}^{\prime}$ remains unchanged from the original best response with $\textbf{\rewardb}$.
Denote by $\mathcal{S}$ the set of all possible random strings and by $\IncludeF_{S'}^*$ the original best response with random string $S'$.
We modify the deviation bribe function $\brbaction{2}{\prime}$ by distributing $R_1^*/3$ evenly across all possible random string $S' \in \mathcal{S}, S' \neq S^*$.
Then, with each random string $S' \neq S^*$, the bribe increases by a small amount of $R_1^*/(3 \cdot (|\mathcal{S}| - 1))$ only for the original best response $\IncludeF_{S'}^*$:
\begin{multline}\label{eq:impossibility:final_dev_bribe}
    \brbaction{2}{\prime}(\IncludeF, S'\neq S^*) =\\
    \begin{cases}
    \rewardb_{P2}(\IncludeF, S') + \frac{R_1^*}{(3 \cdot (|\mathcal{S}| - 1))} & \text{if } \IncludeF = \IncludeF_{S'}^*; \\
    \rewardb_{P2}(\IncludeF,S') & \text{otherwise.}
    \end{cases}
\end{multline}

Then the original best response $\IncludeF_{S'}^*$ becomes the unique best response to the modified deviation bribe function $\brbaction{2}{\prime}$ and the random string $S'$.

(ii) \textbf{Publisher $\prover_2$'s Revenue under Deviation}.

We first calculate publisher $\prover_2$'s revenue with random string $S^*$ in the deviation case and then calculate her expected revenue over all possible random strings.

Recall that, with the random string $S^*$, the difference between the validator's original best response $\IncludeF^*$ and the new best response $\IncludeF^{\prime}$ is that $\IncludeF^{\prime}$ replaces $\prover_1$'s report~$\report_{P1}$ in $\IncludeF^*$ with $\prover_2$'s report~$\report_{P2}$.
Therefore, due to the symmetric reward property, the reward to publisher~$\prover_2$ from $\IncludeF^{\prime}$ is the same as the reward to publisher~$\prover_1$ from~$\IncludeF^*$:
\begin{equation}\label{eq:impossibility:eq_rev}
R(\IncludeF^{\prime}, \{\report_{P2}\}, S^*) = R(\IncludeF^*, \{\report_{P1}\}, S^*) \,\,.
\end{equation}

The revenue $R_1^*$ of publisher $\prover_1$ with the original best response $\IncludeF^*$ is the total reward from her included reports minus her bribe paid to the validator:
\begin{multline}\label{eq:impossibility:u1}
R_1^* = R_{P1}(\{\report_{P1}\},\rewardb_{P1}, \IncludeF^*, S^*) \\
 = R(\IncludeF^*, \{\report_{P1}\}, S^*) - \rewardb_{P1}(\IncludeF^*, S^*) \,\,.
\end{multline}

With random string $S^*$ and  bribe function $\brbaction{2}{\prime}$ the revenue of publisher $\prover_2$ due to the new best response $\IncludeF^{\prime}$ of the validator is:
\begin{align}\label{eq:impossibility:rev_dev_s}
& R_{P2}(\{\report_{P2}\},\brbaction{2}{\prime}, \IncludeF^{\prime}, S^*) \nonumber\\
\stackrel{Eq. \ref{eq:impossibility:pub_rev}}{=}& R(\IncludeF^{\prime}, \{\report_{P2}\}, S^*) - \brbaction{2}{\prime}(\IncludeF^{\prime}, S^*) \nonumber\\
\stackrel{Eq. \ref{eq:impossibility:dev_bribe}}{=}& R(\IncludeF^{\prime}, \{\report_{P2}\}, S^*) - \nonumber\\ & (\rewardb_{P2}(\IncludeF^*,S^*) +  \rewardb_{P1}(\IncludeF^*,S^*) + R_1^*/3) \nonumber\\
\stackrel{Eq. \ref{eq:impossibility:eq_rev}}{=}& (R(\IncludeF^*, \{\report_{P1}\}, S^*) - \rewardb_{P1}(\IncludeF^*, S^*)) - \nonumber\\ & (\rewardb_{P2}(\IncludeF^*,S^*) + R_1^*/3) \nonumber\\
=& R_1^* - \rewardb_{P2}(\IncludeF^*,S^*) - R_1^*/3 \nonumber\\
\stackrel{Eq. \ref{eq:impossibility:pj_zero}}{=}& +  (R(\IncludeF^*, \{\report_{P2}\}, S^*) - \rewardb_{P2}(\IncludeF^*,S^*)) + \nonumber\\ 
& \frac{2}{3} R_1^* \nonumber\\
\stackrel{Eq. \ref{eq:impossibility:pub_rev}}{=} & \frac{2}{3} R_1^* + R_{P2}(\{\report_{P2}\},\rewardb_{P2}, \IncludeF^*, S^*)  \,\,.
\end{align}

With a random string $S' \neq S^*$, publisher $\prover_2$'s revenue from the deviation bribe function $\brbaction{2}{\prime}$ is a bit lower than her revenue from the original bribe function $\rewardb_{P2}$ due to the additional bribe (Equation~\ref{eq:impossibility:final_dev_bribe}).
Recall that the best responses of the validator $\IncludeF_{S'}^*$ in these two cases are identical.
We have
\begin{multline}\label{eq:impossibility:rev_dev_others}
    R_{P2}(\{\report_{P2}\},\brbaction{2}{\prime}, \IncludeF_{S'}^*, S') = \\
    R_{P2}(\{\report_{P2}\},\rewardb_{P2}, \IncludeF_{S'}^*, S') - R_1^*/(3 \cdot (|\mathcal{S}| - 1)) .
\end{multline}

Denote by $\valbr(\cdot)$ the validator's best response.
The expected revenue for publisher $\prover_2$ with the deviation bribe function $\brbaction{2}{\prime}$ is higher than her expected revenue with the original bribe function $\rewardb_{P2}$, when the validator takes the best responses:
\begin{align*}
&E_{S}[R_{P2}(\{\report_{P2}\},\brbaction{2}{\prime}, \valbr(\cdot), S)] \\
=& \frac{1}{|\mathcal{S}|} \left[\sum_{S} R_{P2}(\{\report_{P2}\},\brbaction{2}{\prime}, \IncludeF_{S}^*, S) \right] \\
=& \frac{1}{|\mathcal{S}|} \left[ \substack{R_{P2}(\{\report_{P2}\},\brbaction{2}{\prime}, \IncludeF^{\prime}, S^*) + \\ \sum_{S' \neq S^*} R_{P2}(\{\report_{P2}\},\brbaction{2}{\prime}, \IncludeF_{S'}^*, S')}
 \right] \\
\stackrel{Eq. \ref{eq:impossibility:rev_dev_s}}{=}& \frac{1}{|\mathcal{S}|} \left[ \substack{\frac{2}{3} R_1^* + R_{P2}(\{\report_{P2}\},\rewardb_{P2}, \IncludeF^*, S^*) + \\ \sum_{S' \neq S^*} R_{P2}(\{\report_{P2}\},\brbaction{2}{\prime}, \IncludeF_{S'}^*, S')} \right] \\
\stackrel{Eq. \ref{eq:impossibility:rev_dev_others}}{=}& \frac{1}{|\mathcal{S}|} \left[ \substack{\frac{2}{3} R_1^* + R_{P2}(\{\report_{P2}\},\rewardb_{P2}, \IncludeF^*, S^*) + \\ \sum_{S' \neq S^*} \left(R_{P2}(\{\report_{P2}\},\rewardb_{P2}, \IncludeF_{S'}^*, S') - \frac{R_1^*}{3 \cdot (|\mathcal{S}| - 1)}\right)} \right] \\
=& \frac{1}{|\mathcal{S}|} \left[ \sum_{S} R_{P2}(\{\report_{P2}\},\rewardb_{P2}, \IncludeF_{S}^*, S)  \right]  +  \frac{R_1^*}{3|\mathcal{S}|}\\
=& E_{S}[R_{P2}(\{\report_{P2}\},\rewardb_{P2}, \valbr(\cdot), S)] +  \frac{R_1^*}{3|\mathcal{S}|} \,\,. 
\end{align*}

The above inequality shows that publisher $\prover_2$ can increase her expected revenue by deviating to the new bribe function $\brbaction{2}{\prime}$.
This contradicts the assumption that the protocol is incentive-compatible.

(iii) \textbf{Conclusion}. 

In all, we have shown that for any situation where publishers have positive expected revenue, there exists a profitable deviation for some publisher, demonstrating that the protocol is not incentive-compatible.
This completes the proof of Theorem~\ref{Theorem:impossibility}.
\end{maybeProof}

        \section{Personal Random Rewards for Reporting}\label{sec:mechanism}
    
To overcome the impossibility of symmetric designs, we create controlled asymmetry among participants~(\S\ref{sec:mechanism:easa}).
Building on this concept, we propose a new reporting protocol, Prrr, which employs a second-price-style reward allocation rule~(\S\ref{sec:mechanism:protocol}) and satisfies our desired goals when all participants adhere to the protocol~(\S\ref{sec:mechanism:assessment}).
We analyze the conditions to ensure incentive compatibility~(\S\ref{sec:mechanism:iccondition}) and carefully design the random-value function, providing two specific instantiations, to ensure these incentive-compatibility conditions~(\S\ref{sec:mechanism:valuefunctions}).

            \subsection{Ex-Ante Synthetic Asymmetry (EASA)}\label{sec:mechanism:easa}

As demonstrated in Theorem~\ref{Theorem:impossibility}, symmetric reporting protocols cannot achieve our desired goals. 
To overcome this limitation, we propose \emph{Ex-Ante Synthetic Asymmetry}~(EASA), a novel mechanism-design concept that introduces controlled asymmetry among participants. 

In traditional mechanism design, asymmetry among participants arises naturally due to differences in their profiles. 
However, in our reporting setting, all reports are functionally identical, and publishers operate in a permissionless environment. 
We cannot differentiate between publishers based on the number of reports they generate, as this would compromise fairness and incentivize centralization or identity splitting.

Therefore, EASA assigns a random value to each report using a random-value function $\rewardr$. 
Given a random string~$S$, the random value of a report~$\report$ is determined as~$\rewardr(\report, S)$. 
For a specific random string $S$, the random values of different reports may vary. 
However, if the random string $S$ is unbiased, all reports remain functionally identical in expectation.

            \subsection{The Prrr Protocol}\label{sec:mechanism:protocol}

Based on the EASA concept, we design \emph{Prrr}: \emph{personal random rewards for reporting}.
Prrr achieves incentive compatibility by implementing a reward mechanism rule reminiscent of a second-price auction (\S\ref{sec:mechanism:intuition}).
The Prrr protocol progresses in steps (\S\ref{sec:mechanism:progress}), each consisting of three phases corresponding to the reporting problem (\S\ref{sec:model:report}): publication (\S\ref{sec:mechanism:publication}), inclusion (\S\ref{sec:mechanism:inclusion}), and processing (\S\ref{sec:mechanism:reward}).

\subsubsection{Design Intuition}\label{sec:mechanism:intuition}

EASA assigns a random value to each report, breaking the symmetry among reports.
The key challenge is to design a mechanism that achieves our goals and ensures incentive compatibility, so all participants follow the protocol.

For intuition, consider first a mechanism where all report values are public, and the report included on chain receives a reward equal to its value.
The creator of the highest-value report would bribe the validator with an amount equal to the second-highest value; otherwise, the publisher of the second-highest-value report could out-bribe her, receive the reward, and earn a positive payoff.
However, this mechanism requires publishers to know the values of competing reports, whereas these values are private and off-chain.

Our goal is a truthful mechanism under which publishers are incentivized to publish all reports without bribery, not relying on the knowledge of other reports.
To achieve this, we make the above payoff structure explicit in the protocol.
Specifically, we require the validator to include the two highest-value reports.
The validator receives the second-highest value as her reward, and the publisher of the highest-value report receives the difference between the highest and second-highest values.
This design closely mirrors a second-price auction~\cite{vickrey1961counterspeculation}: the publisher with the highest-value report receives a reward equal to the difference between their report's value and the second-highest, while the validator receives the second-highest value.

\begin{figure}[t]
    \centering
    \includegraphics[width=0.8\linewidth]{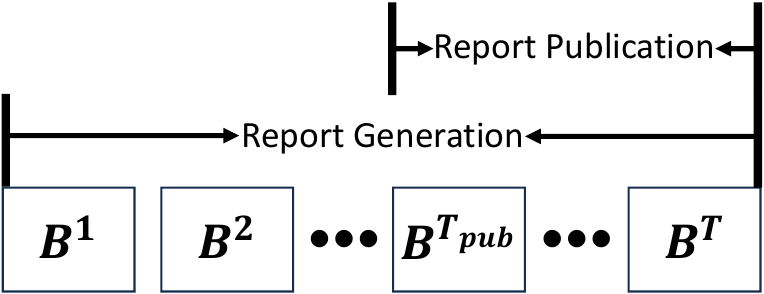}
    \caption{The progress of Prrr in an epoch.}
    \label{fig:progress}
\end{figure}
\subsubsection{Prrr Progress}\label{sec:mechanism:progress}
Assume that an epoch begins at step~$1$ and spans $T$ steps. 
The protocol is parameterized by a publication period $T_{\textit{Pub}}$ (Figure~\ref{fig:progress}).

\begin{itemize}
    \item \textbf{Report-Generation Window:} Publishers may begin generating reports starting from step $1$.
    We can enforce this by requiring reports to include the block hash of $\Includedtx^{1}$, which is only available after the creation of this block.
    \item \textbf{Report-Publication Window:} The smart contract will only process published reports starting from step $T_{\textit{Pub}}$.
    Once the protocol reaches the end of the epoch (after step $T$) or the smart contract receives any report, this window closes, and participants should not take any further actions.
\end{itemize}

This timeline provides publishers with adequate time to prepare reports before competing for inclusion.
Otherwise, if one publisher can generate reports faster than others, they could publish their reports first and secure the reward, leading to centralization.

During the report-publication window, in each step, the protocol executes the report publication, inclusion, and reward-distribution phases sequentially.

Additionally, in the PoS-pattern report (\S\ref{sec:model:participants}), Prrr does not allow publishers to increase their stakes to raise the number of reports they can generate during the report-publication window.

\subsubsection{Report Publication}\label{sec:mechanism:publication}

Prrr requires publishers to publish all their generated reports without bids or bribes~(Algorithm~\ref{alg:reportpublication}).
Specifically, for any publisher~$\prover_j$, denote by~$\textbf{\report}_j$ the set of reports she has generated.
Publisher~$\prover_j$ should publish a set of transactions $\pregen$, where each transaction contains one report from $\textbf{\report}_j$ and a zero transaction fee bid.
Additionally, she should publish a bribe function that assigns zero to all possible outcomes, denoted by $\textbf{0}$.
Notably, the publisher does not need to be aware of other publishers to execute the above protocol.

\begin{algorithm}[t]
    \caption{Report Publication by Publisher~$\prover_j$}
    \label{alg:reportpublication}
    \begin{algorithmic}[1]
    \State \textbf{Input:} Set of reports the publisher~$\prover_j$ generates:~$\textbf{\report}_j$
    \State $\pregen \gets \{ (\report, 0) | \report \in \textbf{\report}_j \}$
    \Statex \Comment{Each transaction comprises a report and a zero bid.}
    \State \textbf{Output:} ($\pregen$, $\textbf{0}$)
    \end{algorithmic}
\end{algorithm}

\subsubsection{Report Inclusion}\label{sec:mechanism:inclusion}

In any step $k$ during the report-publication window, following the report-publication phase, Prrr requires validator $V^k$ to generate the random string~$S^k$, and decide which reports to include in block $\Includedtx^k$.

\subsubsection*{Rationale for Random String Generation}

Prrr uses the random string~$S^k$ to assign random values to reports.
If the randomness was public to both publishers and to the validator, publishers could compute the random values of their reports before publication.
In the second-price-style reward allocation mechanism, only the highest-value report receives a positive reward; all other reports increase the second-highest value and thus reduce the winner's reward.
Consequently, a publisher could strategically publish only her highest-value report and withhold the rest, since publishing additional low-value reports does not increase her reward but could decrease it by raising the second-highest value.
This creates an unfair advantage for publishers generating more reports: They can withhold more reports to further increase their expected rewards compared to publishers generating fewer reports.

To prevent this strategic withholding and ensure fairness, publishers should not know the random string $S^k$ at the time of publication.
Nevertheless, validators should know~$S^k$ when deciding which reports to include in the block, as the the inclusion rule will depend on the random values of the reports (\S\ref{sec:mechanism:intuition}).
To achieve this, Prrr requires each validator~$V^k$ to use Verifiable Random Functions~(VRFs)~\cite{micali1999verifiable} to generate the private random string $S^k$ for each block $\Includedtx^k$.
Anyone can verify its integrity after publication.

\subsubsection*{VRF Secrecy Assumption and Leakage}
Our analysis assumes that validators do not reveal the VRF output $S^k$ before block publication.
While VRFs ensure unpredictability to external observers, validators may still disclose $S^k$ off-chain to colluding publishers.
However, our game-theoretic analysis (\S\ref{sec:ic:robustness}) shows that such premature disclosure does not increase their joint expected revenue.
Thus, correctness does not rely on strict secrecy of $S^k$.

\subsubsection*{Alternative Randomness Mechanisms}

Prrr is also compatible with public randomness mechanisms, such as on-chain randomness beacons~\cite{das2022spurt,bhat2021randpiper} and external oracles providing publicly verifiable randomness~\cite{breidenbach2021chainlink}, which eliminate the need to trust validators not to reveal $S^k$.
However, as we just noted, publishers should not know the realized~$S^k$ when publishing their reports, and the validator should know~$S^k$ when including reports in the block.
Therefore, public randomness would be compatible only if the protocol preserves temporal separation between report publication and randomness revelation, potentially via commit-then-reveal~\cite{choi2023bicorn} or delayed randomness~\cite{boneh2018verifiable}.

\subsubsection*{Concrete Inclusion Process}

Prrr requires validators to generate VRFs based on the external random beacon pulse~$Q^k$.
The VRF consists of two functions~\cite{micali1999verifiable}:

\begin{itemize}
    \item Generation function $\VRFGEN(Q^k, \sk^k) \to S^k$: Validator $V^k$ uses their private key $\sk^k$ and the public beacon $Q^k$ to generate a private random string $S^k$. 
    This string is used to calculate the personal rewards for the reports in block $\Includedtx^k$ and should only be revealed when the block is published.
    \item Verification~function $\VRFVER(Q^k, S^k, \pk^k) \to \{\text{true}, \text{false}\}$: After publication, anyone can use the validator's public key $\pk^k$ to verify that the private string $S^k$ was correctly generated from the public beacon $Q^k$.
\end{itemize}

Prrr adopts a second-price-style report inclusion rule, requiring the validator to include two reports in the block:
The first report is the one with the highest random value, designated as the winning report;
The second report is the one with the second-highest random value, which we will use to implement the reward allocation mechanism in the next phase. 
To further enhance efficiency, Prrr requires the validator to include the second-place report only when its value is strictly higher than the minimum possible $\rewardr$ value~$\minr$, as the smart contract can infer this value during reward allocation without explicit inclusion.

In this phase, validator $V^k$ first generates the secret random string $S^k$ using the VRF generation function.
Next, she collects all published reports from publishers and computes their random values using a random-value function~$\rewardr$.
We denote by $\gentx_j$ the set of transactions published by publisher~$j$.
Then $\cup_{j=1}^n \gentx_j$ is the set of all published transactions from all publishers.
Finally, she puts the two reports with the highest random values into the block according to the second-price-style rule.
The detailed procedure is in Algorithm~\ref{alg:blockgeneration}.

Note that the validator does not need to know publishers and only needs to act according to the transactions she receives.

\begin{algorithm}[t]
    \caption{Report Inclusion by Validator~$V^k$}
    \label{alg:blockgeneration}
    \begin{algorithmic}[1]
    \State \textbf{Input:} Transaction set from all publishers~$\cup_{j=1}^n \textbf{\gentx}_j$, validator's private key $\sk^k$, block index $k$, external random source $Q^k$, and minimum value threshold $\minr$.
    \Function{Sort}{$\cup_{j=1}^n \textbf{\gentx}_j$, $S^k$}
        \State Sort $(\report, \bid)$ in descending order of $\rewardr(\report, S^k)$
        \Statex \Comment{Rank reports by their random values.}
        \State Break ties by ascending order of~$H(\report)$
        \Statex \Comment{Deterministic tie-breaking by hash.}
        \State \textbf{Return:} Sorted list $\sortgentx$
    \EndFunction
    \State Generate the random string $S^k \gets \VRFGEN(Q^k, \sk^k)$
    \State $\sortgentx \gets \Call{Sort}{\cup_{j=1}^n \textbf{\gentx}_j, S^k}$
    \Statex \Comment{Sort all published reports by random value.}
    \State $\preinc^k \gets ()$ \Comment{Initialize empty inclusion list.}
    \If{$\sortgentx \neq \emptyset$}
        \State $\preinc^k.\textit{append}(\sortgentx[1])$
        \Statex \Comment{Include the highest-value report.}
    \EndIf
    
    \If{$|\sortgentx| > 1$}
        \State $(\report, b) \gets \sortgentx[2] $
        \If{$\rewardr(\report, S^k) > \minr$}
            \Statex \Comment{Skip if value equals minimum.}
            \State $\preinc^k.\textit{append}((\report, b))$
            \Statex \Comment{Include the second-highest report.}
        \EndIf
    \EndIf
    \State \textbf{Output:} $\Includedtx^k \gets (k, Q^k, S^k, \preinc^k)$
    \end{algorithmic}
\end{algorithm}

\subsubsection{Report Processing}\label{sec:mechanism:reward}
In any step $k$ during the report-publication window, after the report-inclusion phase, Prrr requires the smart contract to process transactions in block $\Includedtx^k = (k, Q^k, S^k, \IncludeF^k)$ (Algorithm~\ref{alg:rewardallocation}).
The smart contract then allocates rewards based on the second-price-style reward allocation intuition (\S\ref{sec:mechanism:intuition}).

First, the smart contract verifies the integrity of the random string $S^k$ using the VRF verification function.
If the verification fails or the vector of transactions $\IncludeF^k$ is empty, the smart contract takes no further action.
Otherwise, the smart contract proceeds to allocate rewards to publishers and the validator using a second-price-style reward allocation mechanism.
The validator should include the highest and second-highest value reports (if it does not have the minimum value), but in practice this mechanism deals with three scenarios separately.

The first scenario is the \emph{standard case}, which occurs when a block contains exactly two reports and the first,~$\report_1$, has a value higher than or equal to the second,~$\report_2$. 
In addition, the second report's value must exceed the minimum value threshold, $\minr$.
The publisher of the winning $\report_1$ receives the difference between its value and the second-highest value. 
The publisher of the second report $\report_2$ does not receive a reward.
The validator's reward is the value of the second report.
Table~\ref{tab:rewardexamples} provides an example in Case 1: A correctly ordered block with two reports results in the validator earning $R=8$ (the second report's value), while the winning publisher earns the surplus, $10-8=2$.

The second scenario is the \emph{succinct case}, which occurs when a block contains only one report.
In this case, the publisher of the report receives the difference (maybe $0$) between its value and the minimum value, $\minr$.
The validator's reward is the minimum value, $\minr$.
Table~\ref{tab:rewardexamples} illustrates this in Case 2: A block with a single report results in the validator earning $R=2$ (the minimum value), while the publisher earns $10-2=8$.

In all other situations, the mechanism enters a \emph{deviation case}. 
This occurs if the block contains more than two reports (see Table~\ref{tab:rewardexamples}, Case 3), or if it contains two reports but the second report's value is not lower than the first (see Case 4), or if the second report's value equals the minimum value $\minr$ (see Case 5).
In these cases, the publisher of the first report receives the difference between its value and the minimum value $\minr$, while the validator does not receive any reward.

Once the reward allocation is complete, the reporting is finished and the report-publication window ends.

\begin{algorithm}[t]
    \caption{Report Processing by the  Smart Contract}
    \label{alg:rewardallocation}
    \begin{algorithmic}[1]
    \State \textbf{Input:} Block $\Includedtx^k = (k, Q^k, S^k, \IncludeF^k)$, validator's public key $\pk^k$
    \State \textbf{Output:} Publisher rewards $\textbf{R}_{\mathcal{P}}$, Validator reward $R_{\mathcal{V}}$
    \State Initialize $\textbf{R}_{\mathcal{P}} \gets \textbf{0}$, $R_{\mathcal{V}} \gets 0$
    \If{$\IncludeF^k=()$ or $\VRFVER(Q^k, S^k, \pk^k) = \text{false}$} \textbf{return} $\textbf{R}_{\mathcal{P}}, R_{\mathcal{V}}$ 
    \Statex \Comment{Reject if empty block or VRF verification fails.}
    \EndIf
    \State Note that $\IncludeF^k = ((\report_1, \cdot), (\report_2, \cdot), \cdots)$
    \State Let $\prover_{j*}$ be the publisher owns $\report_1$
    \Statex \Comment{Identify the first report's publisher.}
    \State $r_1 \gets \rewardr(\report_1, S^k)$
    \Statex \Comment{Compute the first report's random value.}

    \If{$|\IncludeF^k| = 1$}
        \Statex \Comment{Succinct case: only one report included.}
        \State $R_{\mathcal{V}} \gets \minr$ \Comment{Validator gets the minimum value.}
        \State $\textbf{R}_{\mathcal{P}}[j^*] \gets r_1 - \minr$
        \Statex \Comment{Publisher gets the surplus above minimum.}
    \ElsIf{$|\IncludeF^k| = 2$}
        \State $r_2 \gets \rewardr(\report_2, S^k)$
        \Statex \Comment{Compute the second report's random value.}
        \If{$r_1 \ge r_2$ \textbf{and} $r_2 > \minr$}
            \Statex \Comment{Standard case: reports correctly ordered.}
            \State $R_{\mathcal{V}} \gets r_2$
            \Statex \Comment{Validator gets the second-price value.}
            \State $\textbf{R}_{\mathcal{P}}[j^*] \gets r_1 - r_2$
            \Statex \Comment{Publisher gets the difference.}
        \Else
        \Statex \Comment{Deviation case: wrong order or minimum value.}
            \State $\textbf{R}_{\mathcal{P}}[j^*] \gets r_1 - \minr$
            \Statex \Comment{Validator gets no reward under the deviation case.}
        \EndIf
    \Else \Comment{Deviation case: more than two reports included.}
        \State $\textbf{R}_{\mathcal{P}}[j^*] \gets r_1 - \minr$
        \Statex \Comment{Validator gets no reward under the deviation case.}
    \EndIf

    \State \textbf{return} $\textbf{R}_{\mathcal{P}}, R_{\mathcal{V}}$
    \end{algorithmic}
\end{algorithm}

\begin{table}[t!]
    \renewcommand{\arraystretch}{1.1}
    \setlength{\tabcolsep}{1.85pt} 
    \scriptsize
    \centering
    \begin{tabular}{|c|c|c|c|c|c|c|c|}
        \hline
        \multirow{2}{*}{} & \multicolumn{3}{|c|}{Block Contents} & \multicolumn{4}{c|}{Reward for} \\ 
        \cline{5-8}
        & \multicolumn{3}{|c|}{($\minr=2$)} & Validator & \cellcolor{blue!30} $\report$ A & \cellcolor{purple!30} $\report$ B & \cellcolor{orange!30} $\report$ C\\ 
        \hline
        \multirow{2}{*}{Case 1} 
            & \cellcolor{blue!30} $\report$ A & \cellcolor{purple!30} $\report$ B &  & \multirow{2}{*}{$8$} & \multirow{2}{*}{$10-8=2$} & \multirow{2}{*}{$0$} & \multirow{2}{*}{$0$} \\
            & \cellcolor{blue!30} $R=10$ & \cellcolor{purple!30} $R=8$ &  &  &  &  &  \\
        \hline
        \multirow{2}{*}{Case 2} 
            & \cellcolor{blue!30} $\report$ A &  &  & \multirow{2}{*}{$2$} & \multirow{2}{*}{$10-2=8$} & \multirow{2}{*}{$0$} & \multirow{2}{*}{$0$} \\
            & \cellcolor{blue!30} $R=10$ &  &  &  &  &  &  \\
        \hline
        \multirow{2}{*}{Case 3} 
            & \cellcolor{blue!30} $\report$ A & \cellcolor{purple!30} $\report$ B & \cellcolor{orange!30} $\report$ C & \multirow{2}{*}{$0$} & \multirow{2}{*}{$10-2=8$} & \multirow{2}{*}{$0$} & \multirow{2}{*}{$0$} \\
            & \cellcolor{blue!30} $R=10$ & \cellcolor{purple!30} $R=8$ & \cellcolor{orange!30} $R=2$ &  &  &  &  \\
        \hline
        \multirow{2}{*}{Case 4} 
            & \cellcolor{purple!30} $\report$ B & \cellcolor{blue!30} $\report$ A &  & \multirow{2}{*}{$0$} & \multirow{2}{*}{$0$} & \multirow{2}{*}{$8-2=6$} & \multirow{2}{*}{$0$} \\
            & \cellcolor{purple!30} $R=8$ & \cellcolor{blue!30} $R=10$ &  &  &  &  &  \\
        \hline
        \multirow{2}{*}{Case 5} 
            & \cellcolor{blue!30} $\report$ A & \cellcolor{orange!30} $\report$ C &  & \multirow{2}{*}{$0$} & \multirow{2}{*}{$10-2=8$} & \multirow{2}{*}{$0$} & \multirow{2}{*}{$0$} \\
            & \cellcolor{blue!30} $R=10$ & \cellcolor{orange!30} $R=2$ &  &  &  &  &  \\
        \hline
    \end{tabular}
    \caption{Revenue calculations in different scenarios.}
    \label{tab:rewardexamples}
\end{table}

\subsubsection*{Note}
In practical systems, three additional considerations arise. 
First, some blockchain protocols require a minimum fee for each transaction. 
Second, smart contracts process transactions sequentially within a block, rather than having a global view of the entire block. 
And third, publishers can specify a validity duration window for each transaction. 
These considerations do not affect our analysis~(\S\ref{appendix:practical}).

\subsection{Correctness}\label{sec:mechanism:assessment}

Prrr solves the reporting problem assuming all participants follow the protocol.

Prrr achieves efficiency as it processes at most two reports per epoch ($C=2$).
Prrr also achieves progress because both publishers and validators act promptly to publish and include reports. 
Furthermore, Prrr achieves
\textbf{fairness} through randomness: while individual outcomes are asymmetric, every report has an equal probability of winning.

Decentralization requires publishers to have positive expected revenue (ignoring the cost of generating the reports).
Prrr requires publishers not to bribe so that revenue is exactly the reward from the smart contract.
We denote by~$\max^{(2)}$ the second-highest value in a set,  by~$\TotalPubR(N)$ the expected total reward for all publishers when there are~$N$ reports in total, and by $E_{\randomstring}$ the expectation over the randomness $\randomstring$.
According to our design, given any set of $N$ reports $\{\report_1, \report_2, \ldots, \report_N\}$, $\TotalPubR(N)$ is the expected difference between the highest random value and the second-highest random value among $N$ reports:
\begin{multline}\label{eq:total_reward_pub}
    \TotalPubR(N) = \\ E_{\randomstring}\left[\max_{1\leq i \leq N} \rewardr(\report_i, S) - \maxtwo{1\leq i \leq N\;\;\;\;\;}\rewardr(\report_i, S)\right]
\end{multline}

If the random-value function is not constant, this expected difference is strictly positive. 
By adhering to the protocol, the expected reward for each report is $\frac{1}{N}\TotalPubR(N)$, as the random value is ex-ante symmetric for all reports. 
Thus, the expected revenue for a publisher generating $N_j$ reports is $\frac{N_j}{N}\TotalPubR(N)$. 
As long as $\TotalPubR(N) > 0$, every publisher obtains a positive expected revenue by following the protocol, thereby ensuring decentralization.

\subsection{Conditions for Incentive Compatibility}\label{sec:mechanism:iccondition}

Prrr's correctness depends on the assumption that all participants follow the protocol.
Here, we analyze how to ensure that following the protocol maximizes each participant's expected revenue, i.e., the protocol is incentive compatible.
Although our reward allocation mechanism disincentivizes bribes that alter the order of reports (\S\ref{sec:mechanism:reward}), we find that we still need two conditions for the random-value function to ensure incentive compatibility: (1) Reward Monotonicity to prevent report withholding; and (2) Skipping Resistance to prevent bribery for skipping steps.

\subsubsection*{Reward Monotonicity}
One challenge for incentive compatibility is that publishers can withhold reports.

The private randomness (\S\ref{sec:mechanism:inclusion}) ensures that publishers cannot withhold reports based on their random values.
Otherwise, a publisher could only publish her highest-value report, reducing the second-highest value if her report is the highest among all published reports.
Therefore, publishers with more reports could disproportionately increase their expected reward per report.
This privacy is essential to fairness and incentive compatibility. 

However, withholdings can still happen if the expected reward for a publisher decreases when she publishes more reports.
If the total reward for all publishers is non-decreasing with the number of reports, then for any publisher, publishing more reports increases both total reward and her share of the total reward, thus increasing her expected reward.
In addition, in the worst case, where there is only one publisher, her expected reward is also non-decreasing with the number of reports she publishes.
We call this property \emph{reward monotonicity}:

\begin{property}[Reward Monotonicity]\label{property:monotonicity}
    For any two numbers of reports $N_1 < N_2$ up to the maximum possible number of published reports $N_{\max}$, the expected total publisher reward of the larger set must be no less than that of the smaller set:
    \begin{equation*}
        \forall N_1 < N_2 \leq N_{\max}: \TotalPubR(N_1) \leq \TotalPubR(N_2) 
    \end{equation*}
\end{property}

\subsubsection*{Skipping Resistance}

Another challenge for incentive compatibility is that publishers might bribe validators to skip certain steps.
This allows them to take the chance with new private randomness in the next step, potentially increasing their expected reward.
Although publishers cannot predict the random string in advance, they can offer bribes for any potential realization of the random string.

Therefore, if they are not satisfied with the current random string, they may bribe the validator to skip the current step and try another random string in the subsequent step.
This is especially true if the validator's reward for honest inclusion is low, while the publisher's potential gain from a new random draw is high. 
To prevent this, the cost of a successful bribe must exceed the expected gain. 
The minimum a rational validator would accept as a bribe is their reward from honest inclusion, which is at least $\minr$. 
We consider the worst-case scenario where all publishers collaborate to bribe the validator.
To make such a bribe unprofitable, we require the total expected reward for all publishers to be strictly less than $\minr$.
We call this property \emph{skipping resistance}:
\begin{property}[Skipping Resistance]\label{property:bribery_resistance}
    For any number of reports $N$ up to the maximum possible number of published reports $N_{\max}$, the expected total publisher reward is strictly less than the minimum possible validator reward, $\minr$:
    \begin{equation*}
       \forall 1 \leq N \leq N_{\max}: \TotalPubR(N) < \minr
    \end{equation*}
\end{property}

\subsection{Random-Value Function Instantiation}\label{sec:mechanism:valuefunctions}

We now identify two random-value functions 
that satisfy Reward Monotonicity (Property~\ref{property:monotonicity}) and Skipping Resistance (Property~\ref{property:bribery_resistance}).

\subsubsection*{Logarithmic Value Function}
Recall that $\minr$ represents the minimum possible output of the random-value function~$\rewardr_{\textit{log}}(\report, S)$.
Assume that the random oracle $H(\cdot)$ produces values uniformly distributed in the interval $[0, 1)$, and let~$\lambda > 0$ be a parameter that controls the distribution of rewards.
Given a report $\report$ and a random string $S$, we define the logarithmic random-value function as:
\begin{equation*}
    \rewardr_{\textit{log}}(\report, S) = \minr - \frac{1}{\lambda}\ln(1 - H(\report || S)) \,\,.
\end{equation*}

\subsubsection*{Polarized Value Function}
The polarized random-value function has only two possible outcomes: $\minr$ and $\maxr$, where $\maxr > \minr$ and the probability of receiving $\maxr$ is small, denoted by $p$.
Given the random oracle $H$, a report~$\report$ and a random string $S$, we define the polarized random-value function $\rewardr_{\textit{polarized}}(\report, S)$ as:
\begin{equation*}
    \rewardr_{\textit{polarized}}(\report, S) = 
    \begin{cases} 
        \maxr & \text{if } H(\report || S) \leq p; \\
        \minr & \text{otherwise.}
    \end{cases}
\end{equation*}

\subsubsection*{Properties and Comparison}

We prove that both the logarithmic value function and the polarized value function satisfy Reward Monotonicity (Property~\ref{property:monotonicity}) and Skipping Resistance (Property~\ref{property:bribery_resistance}) when their parameters are appropriately chosen.
For the polarized value function, parameter selection requires an upper bound~$N_{\max}$ on the total number of reports generated by all publishers, whereas the logarithmic value function can be parameterized independently of system scale.
Additionally, the expected total reward paid by the smart contract is upper bounded by the constant maximal reward $\maxr$ in the polarized value function.
Given the number of published reports~$N$, the expected total reward grows as $O(\log N)$ for the logarithmic value function.
Further details and analysis are in Appendix~\ref{appendix:value_function}.


        \section{Game}\label{sec:game}

Our model and protocol give rise to a sequential game~(\S\ref{sec:game:progress}) played by publishers and validators.
We formalize the utilities of participants~(\S\ref{sec:game:utility}).

            \subsection{Game Progress and Actions}\label{sec:game:progress}

Given a specified random-value function~$\rewardr$ for the Prrr protocol, we denote the game for the protocol by $\mathcal{G}(\rewardr)$ or simply~$\mathcal{G}$ when the context is clear.
The game is sequential and proceeds in discrete steps where each step corresponds to a model step~(\S\ref{sec:model}).
The game starts when the report-publication phase starts at step~$T_{Pub}$ and ends at step~$T$ when the epoch ends (\S\ref{sec:mechanism:progress}).

We assume that publishers cannot generate additional reports since the game starts.
For Proof-of-Work reports, this is because the generation of reports is computationally intensive and time-consuming, and we can set the publication window to be short.
For Proof-of-Stake reports, generating a report is computationally trivial, but we limit the number of reports by the publisher's stake, which cannot be increased after the game starts (\S\ref{sec:mechanism:progress}).

To simplify notation, we count the steps of the game from $1$ to $m = T-T_{Pub} + 1$ (the length of the publication window) and denote the validator of step~$k \in \{1, \dots, m\}$ by~$V^k$.
In each step, publishers publish their reports, and then a validator includes the reports in a block.

We assume all validators follow the protocol to generate the correct random string.
Recall that random string generated by validator~$V^k$ is $S^k$.
An incorrect $S^k$ would result in the reports being ignored (\S\ref{sec:mechanism:reward}).
Rather than generating an incorrect $S^k$, validator~$V^k$ could include an empty set of reports, achieving the same outcome.

As discussed in Section~\ref{sec:impossibility}, we omit bids from the game since bribes can subsume them.
We abstract away transactions and focus solely on reports.

We assume that publishers publish their reports in each step without loss of generality since it is equivalent to their actual operations (\S\ref{sec:impossibility}).

\subsubsection{Publisher Actions}
At the beginning of each step~$k$, publisher~$\prover_j$ publishes a set of reports $
\pubaction{j}{k}$ which is a subset of her generated reports $\textbf{\report}_{Pj}$, and a bribe function~$\brbaction{j}{k}$.

\subsubsection{Validator Actions}
After the publishers act in step~$k$, validator~$V^k$'s action, denoted by $\IncludeF^k$, is to select and publish an ordered vector of reports to include in the block. 
The validator can only choose reports from the set of all reports from publishers in that step (i.e., $ \cup_{1\leq j \leq n} \pubaction{j}{k}$).

Reward allocation in Prrr depends on the validator’s inclusion action $\IncludeF^k = (\IncR^k_{1}, \IncR^k_{2}, \ldots, \IncR^k_{l})$, which yields three possible cases. 
To simplify subsequent analysis, rather than using the original inclusion action, we reformulate the validator’s action.
Our goal is to let the reward of the first report become the difference between the values of the first and second reports in all cases.
We introduce a dummy report which always has minimum value, denoted by~$\report_{\text{dummy}}$, to achieve this unified representation:

\begin{itemize}
    \item The validator includes exactly two reports ${(\IncR^k_{1}, \IncR^k_{2})}$, and the value of $\IncR^k_{1}$ is no less than that of $\IncR^k_{2}$, and the value of $\IncR^k_{2}$ is higher than the minimum value. 
    This case already matches our desired structure and we keep it unchanged.
    \item If the action is an empty vector, we just keep it unchanged.
    \item In all other cases, including (1) the validator includes more than two reports; (2) the validator includes only one report; and (3) the validator includes exactly two reports $(\IncR^k_{1}, \IncR^k_{2}, \ldots)$ but the value of $\IncR^k_{1}$ is less than that of $\IncR^k_{2}$ or the value of $\IncR^k_{2}$ is the minimum value.
    We insert the dummy report after the first report: ${\IncludeF^k = (\IncR^k_{1}, \report_{\text{dummy}}, \IncR^k_{2}, \ldots, \IncR^k_{l})}$. 
\end{itemize}

Note that, after this reformulation, if we insert the second position, all subsequent reports are shifted one position to the right.
For consistency, we mark $\report_{\text{dummy}}$ as~$\IncR^k_{2}$ and mark $\IncR^k_{i}$ where $i \geq2$ as $\IncR^k_{i+1}$.
After the reformulation, if the action $\IncludeF^k$ contains exactly two reports, the validator's reward is the value of the second report in the first two cases; otherwise, the validator's reward is zero.
We assume that publishers provide bribes based on the reformulated inclusion action.

\subsubsection{Actions after Termination}
The game terminates once a validator includes any report. 
For notational convenience in our solution concept, we define the action space such that after the first non-empty inclusion, all subsequent actions are null. 

\subsubsection{Actions Following Prrr Protocol}

We now define the actions that follow the Prrr protocol (\S\ref{sec:mechanism}).

For a validator~$V^k$, following the Prrr protocol means selecting the report with the highest random value as the first included report. 
If the second-highest value among the published reports exceeds the minimum value, the validator includes that report as the second included report; otherwise, the dummy report $\report_{\text{dummy}}$ is included as the second report.
If multiple reports share the same value, Prrr adopts a tie-breaking rule based on the hash to get a total order~(Algorithm~\ref{alg:rewardallocation}). 
We denote this action by $\IncludeF^k = \preinc$.

For a publisher~$\prover_j$, following the Prrr protocol means she publishes all her reports (since we abstract away transactions), i.e., $\pubaction{j}{k} = \textbf{\report}_{Pj}$, and that they do not offer any bribes, i.e., $\brbaction{j}{k} = \textbf{0}$, as specified in Algorithm~\ref{alg:reportpublication}.

\subsection{Utilities}\label{sec:game:utility} 

We now formalize the utilities of validators and publishers in the game.

\subsubsection{Validator Utility}\label{sec:game:revenue:validator}

In step~$k$, for a given action ${\IncludeF^k = (\IncR_{1}^k, \IncR_{2}^k, \ldots)}$ of validator~$V^k$, the bribe functions $\textbf{\rewardb}^k = (\brbaction{1}{k}, \cdots, \brbaction{n}{k})$ from all publishers, and the random string~$S^k$, the revenue of validator~$V^k$ is the reward from the smart contract plus the sum of all bribes.
Since a validator only acts in a single step, that revenue is her utility.
We denote by $\indicator$ the indicator function, that is,~$\indicator(\textit{True}) = 1$ and $\indicator(\textit{False}) = 0$.
The utility of validator~$V^k$ is thus:

\begin{multline}\label{eq:validator:utility}
    \vrev^k(\IncludeF^k = (\IncR_1^k, \IncR_2^k, \ldots),\textbf{\rewardb}^k, S^k) = \\\indicator(|\IncludeF^k|=2) \cdot \rewardr(\IncR_2^k, S^k) + \\\sum_{j=1}^{n} \brbaction{j}{k}(\IncludeF^k, S^k).
\end{multline}

\subsubsection{Publisher Utility} \label{sec:game:revenue:publisher}
The report generation cost of each publisher is constant and independent of their actions in the game, so we omit it from the utility function.

The utility of a publisher is the sum of her revenue across all steps in the epoch.
We first focus on the revenue in a single step.
Given publisher~$\prover_j$'s action $(\pubaction{j}{k}, \brbaction{j}{k})$, the random string $S^k$, and the subsequent action of the validator ${\IncludeF^k = (\IncR_1^k, \IncR_2^k, \ldots)}$, the revenue of publisher~$\prover_j$ in step~$k$ is the revenue from reports (only when included as the first report) minus bribes paid:
\begin{multline}\label{eq:publisher:revenue}
    \pubsinglesteprev{j}{k}(\pubaction{j}{k}, \brbaction{j}{k},\IncludeF^k, S^k) = \\  \indicator(\IncR_1^k \in \pubaction{j}{k}) \cdot (\rewardr(\IncR_1^k, S^k) - \\\rewardr(\IncR_2^k, S^k))  -  \brbaction{j}{k}(\IncludeF^k, S^k) \,\,.
\end{multline}

The utility of publisher~$\prover_j$ over the entire epoch is the expected sum of her revenues $\pubsinglesteprev{j}{k}$ across all steps and random strings.

            \section{Incentive Compatibility of Prrr}\label{sec:ic}
We first analyze the best response of participants in the game~(\S\ref{sec:ic:br}).
We prove that all participants following the protocol is an equilibrium~(\S\ref{sec:ic:eq}) robust against collusion, Sybil attacks and deviations that affect the revenue of other participants~(\S\ref{sec:ic:robustness}).


\subsection{Best Response}\label{sec:ic:br}

We use a standard solution concept for our sequential game, the Subgame Perfect Nash Equilibrium (SPNE).
Denote the subgame from step~$k$ to the last step $m$ by $\mathcal{G}^k$.
In the SPNE, in every step $k$, each participant selects a best response that maximizes their utility in the subgame $\mathcal{G}^k$, assuming all participants will act optimally in subsequent subgames $\mathcal{G}^{k+1}$.


\subsubsection{Validator's Best Response}

Unlike publishers, a validator only acts in a single step, and her utility only depends on the actions of publishers and her own action in that step (Equation~\ref{eq:validator:utility}).
Therefore, her best response is independent of previous and future steps.
For a validator~$V^k$ in step~$k$, her best response $\valbr^k(\textbf{\gentx}^k, \textbf{\rewardb}^k,S^k)$ is simply to maximize her utility (Equation~\ref{eq:validator:utility}) based on the actions of all publishers in that step ${\textbf{\gentx}^k = (\pubaction{1}{k}, \ldots, \pubaction{n}{k})}$ and ${\textbf{\rewardb}^k = (\brbaction{1}{k}, \ldots, \brbaction{n}{k})}$ and the random string~$S^k$. So for all validator actions~$\IncludeF^k$:
\begin{multline*}
    \vrev^k(\valbr^k(\textbf{\gentx}^k, \textbf{\rewardb}^k, S^k), \textbf{\rewardb}^k, S^k) \geq \\ \vrev^k(\IncludeF^k, \textbf{\rewardb}^k, S^k).
\end{multline*}

We denote $\valbr^k(\textbf{\gentx}^k, \textbf{\rewardb}^k,S^k)$ by $\valbr^k(\cdot)$ when the parameters are clear from the context.

\subsubsection{Publisher Best Response}

For a publisher~$\prover_j$, the best response in step~$k$ depends on the actions and random strings in all subsequent steps since she acts and earns revenue in all steps of the game.
Therefore, we use backward induction to analyze the best response.

If no validator has included any report before step~$k$, the game proceeds in step~$k$.
The revenue of validator~$V^k$ in step~$k$ (Equation~\ref{eq:validator:utility}) and the revenue of publisher~$\prover_j$ in step~$k$~(Equation~\ref{eq:publisher:revenue}) depend only on the actions in step~$k$ and the random string $S^k$. 
Denote by $\puballrev{j}(\mathcal{G}^k)$ the expected revenue of publisher~$\prover_j$ in subgame~$\mathcal{G}^k$ when all participants follow the best response strategies.

Given the revenue of the subgame starting from next step~$\puballrev{j}(\mathcal{G}^{k+1})$, the action of publisher~$\prover_j$ in this step~$(\pubaction{j}{k}, \brbaction{j}{k})$, the actions of other publishers in this step $\textbf{\gentx}^k_{-Pj}$ and $\textbf{\rewardb}^k_{-Pj}$, and the best response of the validator~$\valbr^k(\cdot)$ to these actions, the utility of publisher~$\prover_j$ in the subgame~$\mathcal{G}^k$ is the expected revenue in step~$k$ plus the expected revenue in the subsequent subgame if validator~$V^k$ does not include any report (thus the game continues):
\begin{multline}\label{eq:publisher:utility}
    \pubutil{j}^{k \sim m}(\pubaction{j}{k},\brbaction{j}{k}, \textbf{\gentx}^k_{-Pj}, \textbf{\rewardb}^k_{-Pj}) = \\
    E_{S^k} \left[\substack{\pubsinglesteprev{j}{k}(\pubaction{j}{k}, \brbaction{j}{k}, \valbr^k(\cdot), S^k) 
    + \\ \indicator(\valbr^k(\cdot) = ()) \cdot \puballrev{j}(\mathcal{G}^{k+1})}
    \right].      
\end{multline}

Denote by $\pubbr_j^k(\textbf{\gentx}^k_{-Pj}, \textbf{\rewardb}^k_{-Pj},  \puballrev{j}(\mathcal{G}^{k+1}))$ the best response of publisher~$\prover_j$ in step~$k$.
We denote it by~$\pubbr_j(\cdot)$ when the parameters are clear from the context.
For any action $(\pubaction{j}{k},\brbaction{j}{k})$ in step~$k$, the best response strategy has no less utility:
\begin{multline}\label{eq:publisher:bestresponse}
    \pubutil{j}^{k \sim m}(\pubbr_j(\cdot), \textbf{\gentx}^k_{-Pj}, \textbf{\rewardb}^k_{-Pj}) \geq \\
    \pubutil{j}^{k \sim m}(\pubaction{j}{k},\brbaction{j}{k}, \textbf{\gentx}^k_{-Pj}, \textbf{\rewardb}^k_{-Pj}) \,\,.
\end{multline}

\subsection{Prrr Equilibrium}\label{sec:ic:eq}

We show that publishers and validators following the Prrr protocol constitutes an SPNE.
We prove this using backward induction, starting from the final step of the game and determining the best response for each player at each step, assuming all future players will also act optimally.

Recall that the action following the Prrr protocol for all publishers $\prover_j$ is to publish all reports and provide a bribe function that always outputs 0, i.e., $(\textbf{\report}_{Pj}, \textbf{0})$.
And for a validator $V^k$ the desired action $\preinc$ is to include the report with the highest random value as the first report and the report with the second-highest random value as the second report.
At this point, we do not rely on the specific random-value function $\rewardr$ but only require it to satisfy Reward Monotonicity (Property~\ref{property:monotonicity}) and Skipping Resistance~(Property~\ref{property:bribery_resistance}).

\subsubsection{Single-Step Analysis}
To analyze the full game, we first analyze the case of a single step to get some preliminary insights.

We first focus on the case where all publishers do not offer any bribes in step~$k$.
Then, the revenue of validator~$V^k$ (Equation~\ref{eq:validator:utility}) only comes from the reward from the smart contract.
The reward is at most the second-highest random value among all published reports, which is exactly the reward she would get by following the Prrr protocol.
Therefore, following the Prrr protocol is a best response for the validator.

\begin{observation}\label{lemma:validator:nobribery_br}
    In step~$k$, if all publishers do not provide bribes, following the Prrr protocol is a best response for validator~$V^k$.
\end{observation}

For any publisher, according to Equation~\ref{eq:publisher:revenue}, we have:

\begin{observation}\label{lemma:publisher:nobribery_nonnegative}
    If publisher $\prover_j$ does not provide any bribe in step~$k$, her revenue in this step is non-negative.
\end{observation}

For any publisher $\prover_j$ with action $(\pubaction{j}{k}, \brbaction{j}{k})$ in step~$k$, if she provides no bribe (i.e., $\brbaction{j}{k} = \mathbf{0}$), her expected revenue is simply the expected reward from the smart contract (see Equation~\ref{eq:publisher:revenue}). 
When validator $V^k$ follows the Prrr protocol, each report has an equal probability to win the reward. 
Thus, the expected revenue for publisher~$\prover_j$ in this step is the proportion of the number of reports she publishes times the expected total reward $\TotalPubR(\cdot)$ distributed by the smart contract:

\begin{observation}\label{observation:revenue_no_bribery}
    Given any report set $\pubaction{j}{k}$ published by publisher~$\prover_j$ in step~$k$, if she provides no bribe~(i.e.,~$\brbaction{j}{k} = \mathbf{0}$) and validator $V^k$ follows the Prrr protocol, the expected revenue of publisher~$\prover_j$ in this step is:
    \begin{multline}
    E_{S^k}[\pubsinglesteprev{j}{k}(\pubaction{j}{k}, \mathbf{0}, \preinc, S^k)] = \\ \frac{|\pubaction{j}{k}|}{\sum_{1 \leq i \leq n} |\pubaction{i}{k}|} \cdot  \TotalPubR\left(\sum_{1 \leq i \leq n} |\pubaction{i}{k}|\right).
\end{multline}
\end{observation}

If $\prover_j$ publishes all her reports (i.e., $\pubaction{j}{k} = \report_j$), the total reward $\TotalPubR(\cdot)$ is maximized due to Reward Monotonicity (Property~\ref{property:monotonicity}).
The proportion of her reports~$\frac{|\pubaction{j}{k}|}{\sum_{1 \leq i \leq n} |\pubaction{i}{k}|}$ is also maximized.
Therefore, when the validator follows the Prrr protocol, any publisher maximizes her expected revenue by publishing all her reports without providing any bribe.

\begin{observation}\label{observation:publisher:allreports_nobribery}
     If the random-value function satisfies Reward Monotonicity (Property~\ref{property:monotonicity}), then in step~$k$, if all publishers provide no bribe ($\brbaction{i}{k} = \mathbf{0}$ for all $i$) and the validator follows the Prrr protocol, any publisher $\prover_j$ has no less expected revenue by publishing all her reports $\textbf{\report}_j$ than by publishing any subset of her reports $\pubaction{j}{k} \subseteq \textbf{\report}_j$:
    \begin{multline*}
        E_{S^k}[\pubsinglesteprev{j}{k}(\textbf{\report}_j, \mathbf{0}, \preinc, S^k)] \geq \\ E_{S^k}[\pubsinglesteprev{j}{k}(\pubaction{j}{k}, \mathbf{0}, \preinc, S^k)] \,\,.
    \end{multline*}
\end{observation}

Now we turn to the case where publisher~$\prover_j$ provides a bribe in step~$k$ while all other publishers follow the Prrr protocol.
We prove that such bribery does not increase her single-step revenue compared to publishing the same report set without any bribe:
\begin{lemma}\label{lemma:publisher:deviation_revenue_compreto_nobr}
    In step~$k$ with any random string $S^k$, assume that all publishers other than publisher~$\prover_j$ follow the Prrr protocol in step~$k$.
    Then, the revenue of publisher~$\prover_j$ with any action $(\pubaction{j}{k}, \brbaction{j}{k})$ in step~$k$ and any best response~$\valbr^k(\cdot)$ of validator~$V^k$ is no higher than her revenue with the same report set but no bribe, $(\pubaction{j}{k}, \mathbf{0})$, if the validator follows the Prrr protocol:
    \begin{multline}\label{eq:publisher:deviation_revenue_compreto_nobr}
        \pubsinglesteprev{j}{k}(\pubaction{j}{k}, \brbaction{j}{k},\valbr(\cdot), S^k) \leq \\\pubsinglesteprev{j}{k}(\pubaction{j}{k}, \mathbf{0},\preinc, S^k)
    \end{multline}
\end{lemma}

Intuitively, we first calculate the upper bound of the combined revenue of publisher~$\prover_j$ and the utility of validator~$V^k$. 
Next, we determine the lower bound of the utility of validator~$V^k$ with her best response. 
Using these bounds, we derive an upper limit for the revenue of publisher~$\prover_j$ and compare it to her revenue when she offers no bribe.
\begin{showWhenSubmit}The proof is in the full version of this report~\cite{anonymous2025prrr}.\end{showWhenSubmit}

\begin{maybeProof}

    When the publisher~$\prover_j$ takes action $(\pubaction{j}{k}, \textbf{0})$, according to Observation~\ref{lemma:publisher:nobribery_nonnegative}, since there is no bribery from $\prover_j$, her revenue is non-negative.
    If taking action~$(\pubaction{j}{k}, \brbaction{j}{k})$ has higher revenue, its revenue must be positive.
    This is only possible when the validator~$V^k$ includes $\prover_j$'s report as the first report in any of her best response (Equation~\ref{eq:validator:utility}), i.e., $\report_1^k \in \pubaction{j}{k}$.

    Since all other publishers follow the Prrr protocol, which does not provide bribe, the revenue of the validator~$V^k$~(Equation~\ref{eq:validator:utility}) with action $\IncludeF^k = (\report_1^k, \report_2^k, \ldots)$ is:
    \begin{multline*}
        \vrev^k(\IncludeF^k, \textbf{\rewardb}^k, S^k) = \\ \indicator(|\IncludeF^k|=2) \cdot \rewardr(\report_2^k, S^k) + \brbaction{j}{k}(\IncludeF^k, S^k).
    \end{multline*}

    Since $\report_1^k \in \pubaction{j}{k}$, the sum of the revenue of publisher $\prover_j$ and the utility of validator~$V^k$ is:
    \begin{align*}
        &\pubsinglesteprev{j}{k}(\pubaction{j}{k}, \brbaction{j}{k},\IncludeF^k, S^k) + \\
        &\vrev^k(\IncludeF^k, \textbf{\rewardb}^k, S^k) \\ 
        =& \indicator(\report_1^k \in \pubaction{j}{k}) \cdot (\rewardr(\report_1^k, S^k) - \rewardr(\report_2^k, S^k))
         + \\ 
        &\indicator(|\IncludeF^k|=2) \cdot \rewardr(\report_2^k, S^k)\\
       \leq&(\rewardr(\report_1^k, S^k) - \rewardr(\report_2^k, S^k))  + \rewardr(\report_2^k, S^k) \\
        \leq & \rewardr(\report_1^k, S^k) \\
        \leq & \max_{\report \in \pubaction{j}{k}} \rewardr(\report, S^k) \,\,.
    \end{align*}

    By following the Prrr protocol, the utility of the validator~$V^k$ is the second-highest random value among all published reports, i.e., $\max^{(2)}_{\report \in \cup_{i=1}^n \pubaction{i}{k}} \rewardr(\report, S^k)$, plus the bribe from publisher~$\prover_j$:
    \begin{multline*}
        \vrev^k(\preinc, \textbf{\rewardb}^k, S^k)= \\ \maxtwo{\report \in \cup_{i=1}^n \pubaction{i}{k}} \rewardr(\report, S^k) + \brbaction{j}{k}(\preinc, S^k).
    \end{multline*}

    The revenue of the best response of validator~$V^k$ is no less than the revenue when following the Prrr protocol.
    Hence, the revenue of publisher $\prover_j$ with action $(\pubaction{j}{k}, \brbaction{j}{k})$ is bounded by:
    \begin{multline}\label{eq:publisher:deviationwithbriberevenue}
        \pubsinglesteprev{j}{k}(\pubaction{j}{k}, \brbaction{j}{k},\valbr(\cdot), S^k) \leq  \\ \max_{\report \in \pubaction{j}{k}} \rewardr(\report, S^k) -  {\max_{\report \in \cup_{i=1}^n \pubaction{i}{k}}}^{(2)} \rewardr(\report, S^k) - \\ \brbaction{j}{k}(\IncludeF^k, S^k).
    \end{multline}

    We now turn to the case where the publisher $\prover_j$ adopts the action $(\pubaction{j}{k}, \mathbf{0})$ and validator~$V^k$ follows the Prrr protocol.
    If~$\prover_j$'s published report has the highest random value, i.e., $\max_{\report \in \pubaction{j}{k}} \rewardr(\report, S^k) = \max_{\report \in \cup_{i=1}^n \pubaction{i}{k}} \rewardr(\report, S^k)$, her revenue is the difference between the highest and the second-highest random values among all published reports; if not, her revenue is zero:
    \begin{multline}\label{eq:publisher:deviationwithoutbriberevenue}
        \pubsinglesteprev{j}{k}(\pubaction{j}{k}, \mathbf{0},\preinc, S^k) = \\ \indicator\left(\max_{\report \in \pubaction{j}{k}} \rewardr(\report, S^k) = \max_{\report \in \cup_{i=1}^n \pubaction{i}{k}} \rewardr(\report, S^k)\right) \cdot \\ (\max_{\report \in \pubaction{j}{k}} \rewardr(\report, S^k) - \maxtwo{\report \in \cup_{i=1}^n \pubaction{i}{k}} \rewardr(\report, S^k))
    \end{multline}

    If~$\prover_j$'s published report has the highest random value, the revenue with action $(\pubaction{j}{k}, \mathbf{0})$ is (Equation~\ref{eq:publisher:deviationwithoutbriberevenue}) $$\max_{\report \in \pubaction{j}{k}} \rewardr(\report, S^k) - \max^{(2)}_{\report \in \cup_{i=1}^n \pubaction{i}{k}} \rewardr(\report, S^k),$$ which is no less than the revenue with action $(\pubaction{j}{k}, \brbaction{j}{k})$ (Equation~\ref{eq:publisher:deviationwithbriberevenue}).
    Otherwise, if~$\prover_j$'s published report does not have the highest random value, the revenue with action $(\pubaction{j}{k}, \mathbf{0})$ is zero~(Equation~\ref{eq:publisher:deviationwithoutbriberevenue}).
    In this case, the highest random value of reports published by publisher $\prover_j$ is no higher than the second-highest random value among all published reports, i.e., $\max_{\report \in \pubaction{j}{k}} \rewardr(\report, S^k) \leq \max^{(2)}_{\report \in \cup_{i=1}^n \pubaction{i}{k}} \rewardr(\report, S^k)$.
    Therefore, the revenue with action $(\pubaction{j}{k}, \brbaction{j}{k})$ (Equation~\ref{eq:publisher:deviationwithbriberevenue}) is no higher than zero, which is again no higher than the revenue with action $(\pubaction{j}{k}, \mathbf{0})$ (Equation~\ref{eq:publisher:deviationwithoutbriberevenue}):
    \begin{multline*}
        \pubsinglesteprev{j}{k}(\pubaction{j}{k}, \brbaction{j}{k},\valbr(\cdot), S^k) \leq \\ \pubsinglesteprev{j}{k}(\pubaction{j}{k}, \mathbf{0},\preinc, S^k). \qedhere
    \end{multline*}
\end{maybeProof}

\subsubsection{Full Game Equilibrium}

We now consider the SPNE of the full game $\mathcal{G}$.

\begin{theorem}\label{theorem:ic}
    In the game $\mathcal{G}(\rewardr)$, if the random-value function $\rewardr$ satisfies Reward Monotonicity (Property~\ref{property:monotonicity}) and Skipping Resistance (Property~\ref{property:bribery_resistance}), then all participants following the Prrr protocol constitute a Subgame Perfect Nash Equilibrium (SPNE).
\end{theorem}

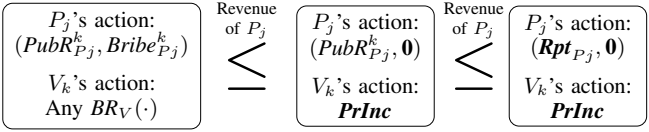
\begin{figure}[t]
    \centering

    \tikzset{
        blockA/.style = {rectangle, rounded corners, draw, align=center,
                         minimum width=0.30\columnwidth, text width=0.30\columnwidth,
                         minimum height=1.1cm, font=\small},
        blockB/.style = {rectangle, rounded corners, draw, align=center,
                         minimum width=0.20\columnwidth, text width=0.20\columnwidth,
                         minimum height=1.1cm, font=\small}
    }

    \resizebox{\columnwidth}{!}{%
    \begin{tikzpicture}[font=\small, node distance=35mm and 8mm]

        \node[blockA] (pub) {$\prover_j$'s action:\\ 
        $(\pubaction{j}{k}, \brbaction{j}{k})$\\[0.6em]
        $V_k$'s action:\\
        Any $\valbr(\cdot)$};

        \node[blockB, right=1.35 of pub] (val) {$\prover_j$'s action:\\ 
        $(\pubaction{j}{k}, \textbf{0})$\\[0.6em]
        $V_k$'s action:\\
        \preinc};

        \node[blockB, right=1.05 of val] (out) {$\prover_j$'s action:\\ 
        $(\textbf{\report}_{Pj}, \textbf{0})$\\[0.6em]
        $V_k$'s action:\\
        \preinc};

        \node at ($(pub)!0.55!(val)$) {\Huge $\leq$};
        \node[font=\scriptsize, align=center] at ($(pub)!0.55!(val)+(0,0.65)$) {Revenue\\of $\prover_j$};

        \node at ($(val)!0.51!(out)$) {\Huge $\leq$};
        \node[font=\scriptsize, align=center] at ($(val)!0.51!(out)+(0,0.65)$) {Revenue\\of $\prover_j$};

    \end{tikzpicture}%
    }

    \caption{Proof sketch of Theorem~\ref{theorem:ic}.}
    \label{fig:proof-flow}
\end{figure}

We prove this theorem by backward induction. 
By Observation~\ref{lemma:validator:nobribery_br}, validators have no incentive to deviate from the Prrr protocol when all publishers follow it, as there are no bribes. 
For any publisher~$\prover_j$, at each step of the backward induction, we first show that her expected revenue from any deviation $(\pubaction{j}{k}, \brbaction{j}{k})$ is no higher than from taking~$(\pubaction{j}{k}, \mathbf{0})$ (Lemma~\ref{lemma:publisher:deviation_revenue_compreto_nobr}). 
Next, we show that publishing all reports without bribes $(\report_j, \mathbf{0})$ yields at least as much expected revenue as any subset $(\pubaction{j}{k}, \mathbf{0})$ (Observation~\ref{observation:publisher:allreports_nobribery}). 
The proof structure is illustrated in Figure~\ref{fig:proof-flow}.
\begin{showWhenSubmit}The proof is in the full version of this report~\cite{anonymous2025prrr}.\end{showWhenSubmit}

\begin{maybeProof}

    By Observation~\ref{lemma:validator:nobribery_br}, validators have no incentive to deviate from the Prrr protocol when all publishers follow it, since there is no bribery. 
    Thus, it suffices to show that no publisher can profitably deviate from the Prrr protocol.
    We prove this by backward induction.
    We first prove that in the subgame $\mathcal{G}^m$, which is just the last step $m$, no publisher can profitably deviate.
    Then, we consider the subgame $\mathcal{G}^k$ starting from step~$k$.
    Assuming that in the subgame $\mathcal{G}^{k+1}$ (starting from the next step), all participants follow the Prrr protocol, we prove that no publisher can profitably deviate in step~$k$ as well.

    (1) Subgame $\mathcal{G}^m$ starting from the last step $m$.
    
    In Lemma~\ref{lemma:publisher:deviation_revenue_compreto_nobr}, we established that bribery cannot increase a publisher's revenue when all other publishers follow the Prrr protocol and the validator adopts any best response.
    It remains to show that publishing fewer reports does not improve a publisher's expected revenue when all other publishers follow the Prrr protocol and the validator follows the protocol as well.
    This follows directly from Observation~\ref{observation:publisher:allreports_nobribery}.

    Assume by contradiction that a publisher $\prover_j$ deviates from the Prrr protocol in step $m$ to any action~$(\pubaction{j}{m}, \brbaction{j}{m})$ while all other publishers follow the Prrr protocol and validator $V^m$ follows any best response.
    According to Equation~\ref{eq:publisher:deviation_revenue_compreto_nobr}, the publisher $\prover_j$'s revenue in this case is not higher than her revenue with action $(\pubaction{j}{m}, \mathbf{0})$, which is the same report set but without bribe, if validator $V^m$ follows the Prrr protocol, for any random string $S^m$.
    Therefore, the following holds for the expected revenue over the distribution of random string~$S^m$:
    \begin{multline}\label{eq:publisher:deviationwithbriberevenue_expectation_laststep}
        E_{S^m} [\pubsinglesteprev{j}{m}(\pubaction{j}{m}, \brbaction{j}{m},\valbr(\cdot), S^m)] \leq \\ E_{S^m} [\pubsinglesteprev{j}{m}(\pubaction{j}{m}, \mathbf{0},\preinc, S^m)] \,\,.
    \end{multline}
    This is the first inequality in Figure~\ref{fig:proof-flow}.

    Now we turn to compare the expected revenue of publisher $\prover_j$ with action $(\pubaction{j}{m}, \mathbf{0})$ and action following the Prrr protocol, i.e., $(\textbf{\report}_{Pj}, \mathbf{0})$, when validator $V^m$ follows the Prrr protocol.
    According to Observation~\ref{observation:publisher:allreports_nobribery}, since the value function satisfies the monotonicity property~(Property~\ref{property:monotonicity}), the expected revenue of publisher $\prover_j$ with action~$(\pubaction{j}{m}, \mathbf{0})$ is no higher than the expected revenue with action $(\textbf{\report}_{Pj}, \mathbf{0})$, when validator $V^m$ follows the Prrr protocol:
    \begin{multline*}
        E_{S^m} [\pubsinglesteprev{j}{m}(\pubaction{j}{m}, \mathbf{0},\preinc, S^m)] \leq \\  E_{S^m} [\pubsinglesteprev{j}{m}(\textbf{\report}_{Pj}, \mathbf{0},\preinc, S^m)]
    \end{multline*}

    Combining Equation~\ref{eq:publisher:deviationwithbriberevenue_expectation_laststep} with the previous analysis, we conclude that for publisher $\prover_j$, if all other publishers follow the Prrr protocol, her expected revenue from any action~${(\pubaction{j}{m}, \brbaction{j}{m})}$, with validator $V^m$ taking any best response, is no higher than her expected revenue from publishing all reports with zero bribe $(\textbf{\report}_{Pj}, \mathbf{0})$ when validator~$V^m$ follows the Prrr protocol:
    \begin{multline*}
        E_{S^m} [\pubsinglesteprev{j}{m}(\pubaction{j}{m}, \brbaction{j}{m},\valbr(\cdot), S^m)] \leq \\ E_{S^m} [\pubsinglesteprev{j}{m}(\textbf{\report}_{Pj}, \mathbf{0},\preinc, S^m)] \,\,.
    \end{multline*}
    This is the second inequality in Figure~\ref{fig:proof-flow}.

    Since there is no subsequent step, the utility of publisher $\prover_j$ in subgame $\mathcal{G}^m$ is exactly her revenue in step $m$ according to Equation~\ref{eq:publisher:utility}:
    \begin{multline*}
            \pubutil{j}^m(\pubaction{j}{m},\brbaction{j}{m}, \textbf{\gentx}^m_{-Pj}, \textbf{\rewardb}^m_{-Pj}) = \\
        E_{S^m} \left[
        \pubsinglesteprev{j}{m}(\pubaction{j}{m}, \brbaction{j}{m}, \valbr^m(\cdot), S^m) 
        \right].     
    \end{multline*}

    Therefore, for publisher $\prover_j$, her utility in subgame $\mathcal{G}^m$ with any action ${(\pubaction{j}{m}, \brbaction{j}{m})}$, considering any best response of the validator~$V^m$, is no higher than her utility with action $(\textbf{\report}_{Pj}, \mathbf{0})$ when validator $V^m$ follows the Prrr protocol.

    (2) Subgame $\mathcal{G}^k$ starting from any step $1 \leq k < m$. 
    
    Assume that all participants follow the Prrr protocol in the subgame $\mathcal{G}^{k+1}$ starting from the next step.
    Unlike the base case, we cannot immediately conclude from Lemma~\ref{lemma:publisher:deviation_revenue_compreto_nobr} that bribery does not increase the publisher's utility, since the utility in $\mathcal{G}^k$ includes the expected revenue in $\mathcal{G}^{k+1}$ if validator $V^k$ chooses not to include any report.
    To address this, we show that if a publisher offers a bribe high enough to induce the validator to skip inclusion, her expected revenue does not increase, due to Skipping Resistance (Property~\ref{property:bribery_resistance}).
    Therefore, as in the base case, providing bribes does not increase the expected revenue of the publisher.
    Similarly, by Observation~\ref{observation:publisher:allreports_nobribery}, publishing fewer reports does not increase the expected revenue of a publisher.

    Assume that any publisher $\prover_j$ deviates from the Prrr protocol in step~$k$ to any action $(\pubaction{j}{k} , \brbaction{j}{k})$.
    We denote by $\pubsinglesteprev{j}{k\sim m}(\pubaction{j}{k}, \brbaction{j}{k},\IncludeF, S^k, \puballrev{j}(\mathcal{G}^{k+1}))$ the revenue of publisher $\prover_j$ in subgame $\mathcal{G}^k$ with action $(\pubaction{j}{k}, \brbaction{j}{k})$ if the validator~$V^k$ takes action $\IncludeF$ with random string $S^k$ and the expected revenue of publisher~$\prover_j$ in subgame $\mathcal{G}^{k+1}$ is $\puballrev{j}(\mathcal{G}^{k+1})$.
    This revenue is
    \begin{multline*}
        \pubsinglesteprev{j}{k\sim m}(\pubaction{j}{k}, \brbaction{j}{k},\IncludeF, S^k, \puballrev{j}(\mathcal{G}^{k+1})) = \\ \pubsinglesteprev{j}{k}(\pubaction{j}{k}, \brbaction{j}{k},\IncludeF, S^k) + \\\indicator\left(\IncludeF=()\right) \cdot \puballrev{j}(\mathcal{G}^{k+1}) \,\,.
    \end{multline*}

    We first prove that, if all other publishers follow the Prrr protocol, the expected revenue of publisher~$\prover_j$ in subgame $\mathcal{G}^k$ with action ${(\pubaction{j}{k}, \brbaction{j}{k})}$ if the validator~$V^k$ follows any best response, is no higher than the expected revenue with action $(\pubaction{j}{k}, \mathbf{0})$ if the validator~$V^k$ follows the Prrr protocol, for any random string $S^k$, i.e., 
    \begin{multline*}
        \pubsinglesteprev{j}{k\sim m}(\pubaction{j}{k}, \brbaction{j}{k},\valbr(\cdot), S^k, \puballrev{j}(\mathcal{G}^{k+1})) \leq \\ \pubsinglesteprev{j}{k\sim m}(\pubaction{j}{k}, \mathbf{0},\preinc, S^k, \puballrev{j}(\mathcal{G}^{k+1})).
    \end{multline*}

    We first analyze the revenue of publisher~$\prover_j$ with action $(\pubaction{j}{k}, \mathbf{0})$ if the validator~$V^k$ follows the Prrr protocol.
    Due to Observation~\ref{lemma:publisher:nobribery_nonnegative}, since all participants follow the Prrr protocol in subgame $\mathcal{G}^{k+1}$ by the induction hypothesis, the expected revenue of publisher $\prover_j$ in subgame $\mathcal{G}^{k+1}$ is non-negative, i.e., $\puballrev{j}(\mathcal{G}^{k+1}) \geq 0$.
    Also due to Observation~\ref{lemma:publisher:nobribery_nonnegative}, the revenue of publisher $\prover_j$ in step $k$ without bribe is non-negative, i.e., $\pubsinglesteprev{j}{k}(\pubaction{j}{k}, \mathbf{0},\preinc, S^k) \geq 0$.
    Therefore, publisher~$\prover_j$'s revenue in subgame $\mathcal{G}^k$ with action~$(\pubaction{j}{k}, \mathbf{0})$ is non-negative:
    \begin{equation}\label{eq:publisher:deviationwithoutbriberevenue_expectation}
        \pubsinglesteprev{j}{k\sim m}(\pubaction{j}{k}, \mathbf{0},\preinc, S^k, \puballrev{j}(\mathcal{G}^{k+1})) \geq 0
    \end{equation}


    We now analyze the revenue of publisher~$\prover_j$ with action $(\pubaction{j}{k}, \brbaction{j}{k})$ if the validator~$V^k$ follows any best response.
    This includes two possible cases: (i) the best response of validator~$V^k$ is to include no report, i.e.,~${\valbr(\cdot) = ()}$; or (ii) the best response of validator~$V^k$ is to include some reports, i.e., $\valbr(\cdot) \neq ()$.
    We prove that in both cases, the revenue of publisher~$\prover_j$ with action~$(\pubaction{j}{k}, \brbaction{j}{k})$ is no higher than the revenue with action $(\pubaction{j}{k}, \mathbf{0})$ if the validator~$V^k$ follows the Prrr protocol.

    (i) If the best response of validator~$V^k$ is to include no report with random string $S^k$, the revenue of publisher~$\prover_j$ is her expected revenue in subgame $\mathcal{G}^{k+1}$ minus the bribe she gives to the validator~$V^k$ for including no report.

    Recall that $\puballrev{j}(\mathcal{G}^{k+1})$ is the expected revenue for publisher $\prover_j$ in the next subgame, which equals her report share times the total expected publisher reward.
    Therefore, the expected revenue of publisher $\prover_j$ in subgame~$\mathcal{G}^{k+1}$ is no higher than the total expected publisher reward $\TotalPubR(\sum_{i} |\textbf{\report}_i|)$.
    By Skipping Resistance (Property~\ref{property:bribery_resistance}), the total expected publisher reward $\TotalPubR(\cdot)$ is at most~$\minr$. Therefore, the expected revenue of publisher $\prover_j$ in subgame~$\mathcal{G}^{k+1}$ is upper bounded by~$\minr$:
    \begin{equation}\label{eq:publisher:followprotocol_expectation_nextstep_upperbound}
        \puballrev{j}(\mathcal{G}^{k+1}) \leq \TotalPubR(\sum_{i} |\textbf{\report}_i|) \leq \minr \,\,.
    \end{equation}

    By following the Prrr protocol, the validator~$V^k$ can get at least~$\minr$ from the smart contract.
    Since including no report is the best response of validator~$V^k$, her revenue with this action must be no less than the revenue by following the Prrr protocol, hence no less than $\minr$.
    When validator~$V^k$ includes no report with random string $S^k$, she can only get revenue from the bribe from publisher $\prover_j$ (Equation~\ref{eq:validator:utility}).
    This is because all other publishers follow the Prrr protocol that does not provide bribe and the protocol does not reward the validator for including no report.
    Therefore, the bribe given by publisher $\prover_j$ on the empty inclusion vector is no less than~$\minr$, i.e., $\brbaction{j}{k}((), S^k) \geq \minr$.

    Recall that the expected revenue of publisher~$\prover_j$ is her expected revenue in subgame $\mathcal{G}^{k+1}$ minus the bribe she gives to the validator~$V^k$.
    Since for publisher~$\prover_j$, the bribe is no less than $\minr$ and the expected revenue in subgame~$\mathcal{G}^{k+1}$ is no higher than $\minr$ (Equation~\ref{eq:publisher:followprotocol_expectation_nextstep_upperbound}), the expected revenue of publisher~$\prover_j$ is non-positive:

    \begin{align*}
        &\pubsinglesteprev{j}{k\sim m}(\pubaction{j}{k}, \brbaction{j}{k},(), S^k, \puballrev{j}(\mathcal{G}^{k+1})) \\
        =&\pubsinglesteprev{j}{k}(\pubaction{j}{k}, \brbaction{j}{k},(), S^k) + \puballrev{j}(\mathcal{G}^{k+1}) \\
        =& - \brbaction{j}{k}((), S^k) +\puballrev{j}(\mathcal{G}^{k+1}) \\
        \leq & - \minr + \puballrev{j}(\mathcal{G}^{k+1}) \\
        \leq & 0 
    \end{align*}
    
    Therefore, the revenue of publisher~$\prover_j$ with action $(\pubaction{j}{k}, \brbaction{j}{k})$ if the validator~$V^k$ does not include any report, is no higher than the revenue with action $(\pubaction{j}{k}, \mathbf{0})$ if the validator~$V^k$ follows the Prrr protocol (Equation~\ref{eq:publisher:deviationwithoutbriberevenue_expectation}):
    \begin{multline*}
        \pubsinglesteprev{j}{k\sim m}(\pubaction{j}{k}, \brbaction{j}{k},(), S^k, \puballrev{j}(\mathcal{G}^{k+1})) \leq \\ \pubsinglesteprev{j}{k\sim m}(\pubaction{j}{k}, \mathbf{0},\preinc, S^k, \puballrev{j}(\mathcal{G}^{k+1})) \,\,.
    \end{multline*}

    (ii) When the best response of validator~$V^k$ to action $(\pubaction{j}{k}, \brbaction{j}{k})$ with random string $S^k$ is not an empty vector, the revenue for publisher~$\prover_j$ is her revenue in step~$k$ only (Equation~\ref{eq:publisher:utility}):
    \begin{multline*}
        \pubsinglesteprev{j}{k\sim m}(\pubaction{j}{k}, \brbaction{j}{k},\valbr(\cdot)\neq (), S^k, \puballrev{j}(\mathcal{G}^{k+1})) \\ =\pubsinglesteprev{j}{k}(\pubaction{j}{k}, \brbaction{j}{k},\valbr(\cdot), S^k)\,\,.
    \end{multline*}

    Now we consider the revenue of publisher~$\prover_j$ with action $(\pubaction{j}{k}, \mathbf{0})$ if the validator~$V^k$ follows the Prrr protocol, i.e.,~$\pubsinglesteprev{j}{k\sim m}(\pubaction{j}{k}, \mathbf{0},\preinc, S^k, \puballrev{j}(\mathcal{G}^{k+1}))$.
    It is no less than her revenue in step~$k$, $\pubsinglesteprev{j}{k}(\pubaction{j}{k}, \mathbf{0},\preinc, S^k)$~(Equation~\ref{eq:publisher:utility}):
    \begin{multline*}
        \pubsinglesteprev{j}{k\sim m}(\pubaction{j}{k}, \mathbf{0},\preinc, S^k, \puballrev{j}(\mathcal{G}^{k+1})) \geq \\ \pubsinglesteprev{j}{k}(\pubaction{j}{k}, \mathbf{0},\preinc, S^k) \,\,.
    \end{multline*}

    Combining with Equation~\ref{eq:publisher:deviation_revenue_compreto_nobr}, we have that the revenue of publisher~$\prover_j$ with action $(\pubaction{j}{k}, \brbaction{j}{k})$ if the validator~$V^k$ follows any best response that is not an empty vector, is no higher than the revenue with action $(\pubaction{j}{k}, \mathbf{0})$ if the validator~$V^k$ follows the Prrr protocol:
    \begin{multline*}
        \pubsinglesteprev{j}{k\sim m}(\pubaction{j}{k}, \brbaction{j}{k},\valbr(\cdot)\neq (), S^k, \puballrev{j}(\mathcal{G}^{k+1})) \\ \leq\pubsinglesteprev{j}{k\sim m}(\pubaction{j}{k}, \mathbf{0},\preinc, S^k, \puballrev{j}(\mathcal{G}^{k+1})) \,\,.
    \end{multline*}

    Combining the cases (i) and (ii), we conclude that for any action ${(\pubaction{j}{k}, \brbaction{j}{k})}$ deviating from the Prrr protocol, the revenue of publisher $\prover_j$ if the validator~$V^k$ follows any best response, is not higher than the revenue with action $(\pubaction{j}{k}, \mathbf{0})$ if the validator~$V^k$ follows the Prrr protocol, for any random string $S^k$:
    \begin{multline}\label{eq:publisher:deviationwithbriberevenue_expectation_noval2}
        E_{S^k} [\pubsinglesteprev{j}{k\sim m}(\pubaction{j}{k}, \brbaction{j}{k},\valbr(\cdot), S^k, \puballrev{j}(\mathcal{G}^{k+1}))] \\ \leq E_{S^k} [\pubsinglesteprev{j}{k\sim m}(\pubaction{j}{k}, \mathbf{0},\preinc, S^k, \puballrev{j}(\mathcal{G}^{k+1}))] \,\,.
    \end{multline}

    This is the first inequality in Figure~\ref{fig:proof-flow}.

    Now we turn to compare the expected revenue of publisher $\prover_j$ taking any action without bribery $(\pubaction{j}{k}, \mathbf{0})$ and following the Prrr protocol, i.e., $(\textbf{\report}_{Pj}, \mathbf{0})$, when validator~$V^k$ follows the Prrr protocol (Second inequality in Figure~\ref{fig:proof-flow}).

    Following the Prrr protocol, the validator action $\preinc^k$ is not an empty vector.
    Hence, the expected revenue of publisher~$\prover_j$ with action $(\pubaction{j}{k}, \mathbf{0})$ in the subgame $\mathcal{G}^k$ only includes the expected revenue in step $k$:
    \begin{multline*}
        \pubsinglesteprev{j}{k\sim m}(\pubaction{j}{k}, \mathbf{0},\preinc, S^k, \puballrev{j}(\mathcal{G}^{k+1})) = \\ \pubsinglesteprev{j}{k}(\pubaction{j}{k}, \mathbf{0},\preinc, S^k) \,\,.
    \end{multline*}
    
    Note that following the Prrr protocol is a special case of action $(\pubaction{j}{k}, \mathbf{0})$ by setting $\pubaction{j}{k} = \textbf{\report}_{Pj}$.
    Therefore, the expected revenue of publisher~$\prover_j$ in subgame $\mathcal{G}^k$ with action $(\pubaction{j}{k}, \mathbf{0})$ is also her expected revenue in step $k$:
    \begin{multline*}
        \pubsinglesteprev{j}{k\sim m}(\textbf{\report}_{Pj}, \mathbf{0},\preinc, S^k, \puballrev{j}(\mathcal{G}^{k+1})) = \\\pubsinglesteprev{j}{k}(\textbf{\report}_{Pj}, \mathbf{0},\preinc, S^k) \,\,.
    \end{multline*}

    Due to Observation~\ref{observation:publisher:allreports_nobribery}, the expected revenue of publisher~$\prover_j$ in this step with action $(\pubaction{j}{k}, \mathbf{0})$ is no higher than the expected revenue by following the Prrr protocol.
    This indicates that the expected revenue of publisher $\prover_j$ in subgame $\mathcal{G}^k$ with action $(\pubaction{j}{k}, \mathbf{0})$ is not higher than the expected utility due to following the Prrr protocol since they are exactly the revenue in step $k$:
    \begin{multline*}
        E_{S^k} [\pubsinglesteprev{j}{k\sim m}(\pubaction{j}{k}, \mathbf{0},\preinc, S^k, \puballrev{j}(\mathcal{G}^{k+1}))] \leq \\ E_{S^k} [\pubsinglesteprev{j}{k\sim m}(\textbf{\report}_{Pj}, \mathbf{0},\preinc, S^k, \puballrev{j}(\mathcal{G}^{k+1}))] \,\,.
    \end{multline*}

    Combined with Equation~\ref{eq:publisher:deviationwithbriberevenue_expectation_noval2}, we have that the expected revenue of publisher $\prover_j$ in subgame $\mathcal{G}^k$ with action ${(\pubaction{j}{k}, \brbaction{j}{k})}$ and any best response of validator~$V^k$ is no higher than the expected revenue if she adopts action $(\pubaction{j}{k}, \mathbf{0})$ and the validator~$V^k$ follows the Prrr protocol:
    \begin{multline*}
        E_{S^k} [\pubsinglesteprev{j}{k\sim m}(\pubaction{j}{k}, \brbaction{j}{k},\valbr(\cdot), S^k, \puballrev{j}(\mathcal{G}^{k+1}))] \\ \leq E_{S^k} [\pubsinglesteprev{j}{k\sim m}(\pubaction{j}{k}, \mathbf{0},\preinc, S^k, \puballrev{j}(\mathcal{G}^{k+1}))] \,\,.
    \end{multline*}

    The utility of publisher $\prover_j$ is exactly her expected revenue in subgame $\mathcal{G}^k$.
    Thus, the above equation shows that the utility of publisher~$\prover_j$ in subgame~$\mathcal{G}^k$, for any deviation $(\pubaction{j}{k}, \brbaction{j}{k})$ and any best response of validator $V^k$, is no higher than her utility from following the Prrr protocol (i.e., $(\pubaction{j}{k}, \mathbf{0})$ with the validator following Prrr).
    Therefore, deviating from the Prrr protocol does not increase the utility of publisher~$\prover_j$ in subgame~$\mathcal{G}^k$ when all other publishers follow Prrr and the validator adopts any best response.
    This indicates that all participants following the Prrr protocol constitutes a SPNE in subgame~$\mathcal{G}^k$.

    By induction, all participants following the Prrr protocol constitutes a SPNE in the entire game.
\end{maybeProof}

            \subsection{Equilibrium Robustness}\label{sec:ic:robustness}
In Section~\ref{sec:ic:eq}, we proved that all participants following the Prrr protocol constitutes an SPNE. 
Now, we analyze the robustness of this equilibrium from two perspectives.

First, while the SPNE ensures that no single participant can increase their revenue by deviating, we also consider the possibility of cooperative deviations involving multiple participants. 
This includes two classic threats to mechanism design~\cite{todo2011false,laffont2000mechanism} and blockchain systems~\cite{roughgarden2021transaction,chung2023foundations}: collusion among multiple existing participants; and Sybil attacks, where a single participant creates multiple identities and coordinates their actions to behave cooperatively.
We prove that collusion or Sybil attacks of publishers (\S~\ref{sec:game:robustness:pubcollusion}) and the collusion between publishers and a validator (\S~\ref{sec:game:robustness:valpubcollusion})
cannot increase their total expected revenue by deviating from the Prrr protocol.

Second, we analyze the stability of our SPNE (\S~\ref{sec:ic:robustness:stability}), focusing on the effects of deviations. 
We prove that validator deviations strictly reduce their revenue if Prrr employs the logarithmic random-value function (\S~\ref{sec:mechanism:valuefunctions}). 
For publishers, we demonstrate that no publisher can deviate in a way that influences other participants' revenue without also reducing her own revenue.

\subsubsection{Publisher Collusion and Sybil Attacks}\label{sec:game:robustness:pubcollusion}

We first demonstrate that Sybil-proofness can be derived from collusion-proofness among publishers. 
In a Sybil attack, a publisher creates multiple fake identities and distributes her reports among these identities.
Then these fake identities coordinate their actions to maximize their total expected revenue.
Recall that, following the Prrr protocol, a publisher's expected revenue is proportional to her share of the total reports. 
If a Sybil attack could increase the attacker's total expected revenue, it would imply that, if these fake identities were independent publishers, they could collude to achieve a combined revenue exceeding the proportional share of their total reports. 
Thus, if collusion among multiple publishers cannot increase their total expected revenue, it also ensures that Sybil attacks are not profitable.

Therefore, we focus on analyzing collusion among multiple publishers.
We found that if a subset of publishers collude in the game~$\mathcal{G}$, we can construct a transformed game~$\mathcal{G'}$. 
In this transformed game, the colluding publishers are represented as a single publisher. 
We map the actions of the colluding publishers in~$\mathcal{G}$ to the actions of this single publisher in~$\mathcal{G'}$.
Then the total revenue of colluding publishers in~$\mathcal{G}$ equals the revenue of this single publisher in~$\mathcal{G'}$, while revenues of all other participants remain unchanged.
\begin{lemma}\label{lemma:collusion:transformation}
    In game $\mathcal{G}$, assuming the set of publishers is $\textbf{\prover}=\{\prover_1, \prover_2, \ldots, \prover_n\}$ and their generated reports in step $k$ are $\textbf{\report}_1^k, \textbf{\report}_2^k, \ldots, \textbf{\report}_n^k,$ respectively.
    Denote by $C \subseteq \{\prover_1, \prover_2, \ldots, \prover_n\}$ the colluding publishers.
    We now construct a new game $\mathcal{G'}$ with a single publisher $\prover_{C}$ representing all colluding publishers:
    \begin{itemize}
        \item The publisher set is $\textbf{\prover'} = \{\prover_{C}\} \cup \textbf{\prover} \setminus C$.
        \item Publisher $\prover_{C}$'s reports in step $k$ are the union of all reports from colluding publishers in step $k$: $\textbf{\report}_{PC}^k = \bigcup_{\prover_j \in C} \textbf{\report}_{Pj}^k$.
    \end{itemize}
    Considering any execution of game $\mathcal{G}$, we map the actions of participants to the new game $\mathcal{G'}$ as follows:
    \begin{itemize}
        \item Given actions of colluding publishers in the original game $\{(\pubaction{j}{k}, \brbaction{j}{k})\}_{\prover_j \in C}$ in any step $k$, publisher $\prover_{C}$'s action in step $k$ in the new game $\mathcal{G'}$ is defined as: (1) The published report is the union set of colluders $\pubaction{C}{k} = \bigcup_{\prover_j \in C} \pubaction{j}{k}$. (2) The output of the bribe function is the sum of all colluders' bribe functions: $\brbaction{C}{k}(\cdot, \cdot) = \sum_{\prover_j \in C} \brbaction{j}{k}(\cdot, \cdot)$.
        \item Publishers other than $\prover_{C}$ keep their reports and actions the same as in the original game $\mathcal{G}$.
    \end{itemize}
    Then, for any step and random string, the revenue of the validator and all non-colluding publishers is identical in both executions, and the total revenue of the colluding publishers in $\mathcal{G}$ equals the revenue of $\prover_{C}$ in $\mathcal{G'}$.
\end{lemma}

This lemma can be directly derived from the definition of publisher and validator revenues (Equations~\ref{eq:validator:utility} and~\ref{eq:publisher:revenue}).
\begin{showWhenSubmit}The proof is in the full version of this report~\cite{anonymous2025prrr}.\end{showWhenSubmit}

\begin{maybeProof}
    With the given transformation, the revenue of the validator (Equation~\ref{eq:validator:utility}) remains unchanged since the total bribe and the set of published reports from colluding publishers is preserved.
    The revenue of non-colluding publishers (Equation~\ref{eq:publisher:revenue}) also remains unchanged since it only depends on their own actions and the validator's action.
    We denote by $\IncludeF^k = (\IncR_1^k, \IncR_2^k, \ldots)$ the inclusion vector chosen by the validator in step $k$.
    The sum of the revenue of colluding publishers in the original game $\mathcal{G}$ in step $k$ with random string $S^k$ is
    \begin{align*}
        &\sum_{\prover_j \in C} \pubsinglesteprev{j}{k}(\pubaction{j}{k}, \brbaction{j}{k},\IncludeF^k (\cdot), S^k) \\ 
        =&\sum_{\prover_j \in C} \left(\substack{\indicator(\IncR_1^k \in \pubaction{j}{k}) \cdot \\ (\rewardr(\IncR_1^k, S^k) - \rewardr(\IncR_2^k, S^k))}\right)  - \\
        &\sum_{\prover_j \in C} \brbaction{j}{k}(\IncludeF^k, S^k) \\
        =&  \left(\substack{\indicator(\IncR_1^k \in \bigcup_{\prover_j \in C} \pubaction{j}{k}) \cdot \\ (\rewardr(\IncR_1^k, S^k) - \rewardr(\IncR_2^k, S^k))}\right)
         - \\
        &\sum_{\prover_j \in C} \brbaction{j}{k}(\IncludeF^k, S^k) \,\,.
    \end{align*}
    This sum is exactly the revenue of publisher $\prover_{C}$ in the new game $\mathcal{G'}$ in step $k$ with random string $S^k$.
\end{maybeProof}

Since in the new game $\mathcal{G'}$, all participants following the Prrr protocol constitutes a SPNE (Theorem~\ref{theorem:ic}), colluding publishers cannot increase their total expected revenue by deviating from the Prrr protocol in the original game~$\mathcal{G}$.
Therefore, Prrr is collusion-resistant among publishers.
Since Sybil attacks can be reduced to collusion among multiple publishers, Prrr is also Sybil-proof.

\begin{corollary}\label{corollary:pubcollude_no_deviation}
    In the SPNE of game $\mathcal{G}$ where all participants follow the Prrr protocol, colluding publishers cannot increase their total expected revenue by deviating from the Prrr protocol and a publisher cannot increase her expected revenue by launching a Sybil attack.
\end{corollary}

\subsubsection{Publisher and Validator Collusion}\label{sec:game:robustness:valpubcollusion}

Now we consider collusion between a subset of publishers and the validator in a single step.

We only focus on the collusion between publishers and the validator in the first step, denoted by~$V^1$.
We do not consider the collusion with a future validator since we assume a large validator set (Section~\ref{sec:model:report}) so publishers do not know the identity of future validators~\cite{kiayias2017ouroboros,daian2019snow}.

Since colluding publishers can be represented as a single publisher (Lemma~\ref{lemma:collusion:transformation}), it suffices to analyze collusion between a single publisher $\prover_j$ and validator $V^1$ in the first step.
Then, we have:

\begin{lemma}\label{lemma:collusion:pubval_no_increase}
In game $\mathcal{G}(\rewardr)$ where the random-value function $\rewardr$ satisfies Reward Monotonicity (Property~\ref{property:monotonicity}) and Skipping Resistance (Property~\ref{property:bribery_resistance}), in the SPNE where all participants follow the Prrr protocol, collusion between any publishers and validator $V^1$ in the first step does not increase their total revenue.
\end{lemma}

Suppose $\prover_j$ learns the private random string $S^1$ before acting. 
She can then publish only her report $\report$ that maximizes $\rewardr(\report, S^1)$, rather than publishing all her reports. 
If $\report$ is the highest among all publishers' reports, this strategy avoids the case where another of her reports is the second-highest, which would reduce her reward.
However, this action decreases the validator's revenue, since she can at most get the random value of the third-highest report instead of the second-highest.
Thus, while the publisher may increase her own reward, the total revenue for the colluding pair (publisher and validator) does not increase compared to following the protocol. 
In other words, collusion between a publisher and the validator in the first step cannot increase their combined revenue.
\begin{showWhenSubmit}The proof is in the full version of this report~\cite{anonymous2025prrr}.\end{showWhenSubmit}

\begin{maybeProof}
    First, we present the revenue structure of the colluding parties.
    Next, we analyze their revenue when they adhere to the Prrr protocol.
    Finally, we calculate their revenue with deviation and compare it to the revenue obtained by following the Prrr protocol.

    (1) Revenue Structure of Colluding Parties:
    
    Since colluding publishers can be represented as a single publisher (Lemma~\ref{lemma:collusion:transformation}), it suffices to analyze collusion between a single publisher $\prover_j$ and validator $V^1$ in the first step.
    If $\prover_j$ learns the private random string $S^1$ before acting, her revenue in the game (Equation~\ref{eq:publisher:utility}) consists of her revenue in step 1 (given $S^1$) plus her expected revenue in the subgame~$\mathcal{G}^2$ if validator $V^1$ does not include any report.
    Recall that $\puballrev{j}(\mathcal{G}^{2})$ is the expected revenue for publisher~$\prover_j$ in the subgame starting from step 2.
    We denote publisher~$\prover_j$'s action by $(\pubaction{j}{1}, \brbaction{j}{1})$ and validator~$V^1$'s action by $\IncludeF^1  = (\IncR_1^1, \IncR_2^1, \ldots)$.
    The revenue of publisher $\prover_j$ (Equation~\ref{eq:publisher:utility}) in the game is:

    \begin{align}\label{eq:collusion:pubval:deviation_revenue}
        &\pubsinglesteprev{j}{1\sim m}(\pubaction{j}{1}, \brbaction{j}{1},\IncludeF^1, S^k, \puballrev{j}(\mathcal{G}^{2}))  \nonumber\\
        =&\pubsinglesteprev{j}{1}(\pubaction{j}{1}, \brbaction{j}{1},\IncludeF^1, S^1) + \nonumber\\
        & \indicator(\IncludeF^1 = ()) \cdot \puballrev{j}(\mathcal{G}^{2})\nonumber\\
        \stackrel{Eq.\ref{eq:publisher:revenue}}{=}& 
        \left( \substack{\indicator({\IncR_1^1 \in \pubaction{j}{1}}) \cdot \\ (\rewardr(\IncR_1^1, S^1)- \rewardr(\IncR_2^1, S^1))}\right) - \brbaction{j}{1}(\IncludeF^1, S^1) + \nonumber\\ 
        & \indicator({\IncludeF^1 = ()}) \cdot \puballrev{j}(\mathcal{G}^{2}) \,\,.
    \end{align}

    Recall that ${\textbf{\gentx}^{1} = (\pubaction{1}{1}, \ldots, \pubaction{n}{1})}$ and that ${\textbf{\rewardb}^{1} = (\brbaction{1}{1}, \ldots, \brbaction{n}{1})}$ are the actions of all publishers in step 1.
    Note that all publishers other than $\prover_j$ follow the Prrr protocol, i.e., ${\pubaction{i}{1} = \textbf{\report}_{Pi}}$ and ${\brbaction{i}{1} = \mathbf{0}}$ for all $\prover_i \neq \prover_j$.
    The revenue of validator $V^1$ (Equation~\ref{eq:validator:utility}) with actions $\textbf{\gentx}^{1}, \textbf{\rewardb}^{1}$ and $\IncludeF^1$ is:
    \begin{align}\label{eq:collusion:val_deviation_revenue}
        &R_V^1(\IncludeF^1,\textbf{\rewardb}^1, S^1) \nonumber\\
       =& \sum_{i} \brbaction{i}{1}(\IncludeF^1, S^1) +  \nonumber \\
       &\indicator({|\IncludeF^1|=2}) \cdot \rewardr(\IncR_2^1, S^1) \nonumber\\
        =& \brbaction{j}{1}(\IncludeF^1, S^1) + \indicator({|\IncludeF^1|=2}) \cdot \rewardr(\IncR_2^1, S^1) \,\,.
    \end{align}

    Denote by $R_{\textit{collude}}$ the total revenue of publisher $\prover_j$ and validator $V^1$ (Equation~\ref{eq:validator:utility}) due to the above actions:
    \begin{align}\label{eq:collude_revenue_structure}
        &R_{\textit{collude}} (\textbf{\gentx}^{1}, \textbf{\rewardb}^{1},\IncludeF^1, S^k, \puballrev{j}(\mathcal{G}^{2})) \nonumber \\
         =& \pubsinglesteprev{j}{1\sim m}(\pubaction{j}{1}, \brbaction{j}{1},\IncludeF^1, S^k, \puballrev{j}(\mathcal{G}^{2})) + \nonumber\\
         & R_V^1(\IncludeF^1,\textbf{\rewardb}^1, S^1) \nonumber\\
        \stackrel{Eq. \ref{eq:collusion:pubval:deviation_revenue}}{=}& 
        \left(\substack{\indicator({\IncR_1^1 \in \pubaction{j}{1}}) \cdot\\ (\rewardr(\IncR_1^1, S^1)- \rewardr(\IncR_2^1, S^1))}\right)
         - \brbaction{j}{1}(\IncludeF^1, S^1) +\nonumber\\ 
        & \indicator({\IncludeF^1 = ()}) \cdot \puballrev{j}(\mathcal{G}^{2}) + R_V^1(\IncludeF^1,\textbf{\rewardb}^1, S^1)\nonumber\\
        \stackrel{Eq. \ref{eq:collusion:val_deviation_revenue}}{=}& \left(\substack{\indicator({\IncR_1^1 \in \pubaction{j}{1}}) \cdot\\ (\rewardr(\IncR_1^1, S^1)- \rewardr(\IncR_2^1, S^1))}\right) - \brbaction{j}{1}(\IncludeF^1, S^1) + \nonumber\\ 
        &\indicator({\IncludeF^1 = ()}) \cdot \puballrev{j}(\mathcal{G}^{2}) + \brbaction{j}{1}(\IncludeF^1, S^1) +  \nonumber\\
        &\indicator({|\IncludeF^1|=2}) \cdot \rewardr(\IncR_2^1, S^1) \nonumber\\
        =&\left(\substack{\indicator({\IncR_1^1 \in \pubaction{j}{1}}) \cdot\\ (\rewardr(\IncR_1^1, S^1)- \rewardr(\IncR_2^1, S^1))}\right) + \nonumber\\ 
        & \indicator({\IncludeF^1 = ()}) \cdot \puballrev{j}(\mathcal{G}^{2}) + \nonumber \\
        & \indicator({|\IncludeF^1|=2}) \cdot \rewardr(\IncR_2^1, S^1) \,\,.
    \end{align}

    For notation simplicity, we use ${R_{\textit{collude}} (\pubaction{j}{1}, \brbaction{j}{1},\IncludeF^1, S^k, \puballrev{j}(\mathcal{G}^{2}))}$ instead of ${R_{\textit{collude}} (\textbf{\gentx}^{1}, \textbf{\rewardb}^{1},\IncludeF^1, S^k, \puballrev{j}(\mathcal{G}^{2}))}$, since all other publishers follow the Prrr protocol.

    (2) Revenue When Following the Prrr Protocol:

    When both publisher $\prover_j$ and validator $V^1$ follow the Prrr protocol, the action of validator $V^1$ is ${\preinc = (\IncR_1^1, \IncR_2^1)}$ and the action of publisher $\prover_j$ is $(\textbf{\report}_{Pj},\mathbf{0})$.
    The total revenue of publisher $\prover_j$ and validator~$V^1$ by following the Prrr protocol is:
    \begin{multline*}
        R_{\textit{collude}} (\textbf{\report}_{Pj}, \mathbf{0},\preinc, S^k, \puballrev{j}(\mathcal{G}^{2})) = \\
        \indicator({\IncR_1^1 \in \pubaction{j}{1}}) \cdot (\rewardr(\IncR_1^1, S^1)- \rewardr(\IncR_2^1, S^1))  +  \\\rewardr(\IncR_2^1, S^1) 
    \end{multline*}

    When publisher $\prover_j$ holds the report with the highest value, then $\IncR_1^1 \in \textbf{\report}_{Pj}$, so
    \begin{multline*}
        \indicator({\IncR_1^1 \in \pubaction{j}{1}}) \cdot (\rewardr(\IncR_1^1, S^1)- \rewardr(\IncR_2^1, S^1)) =\\ (\rewardr(\IncR_1^1, S^1)- \rewardr(\IncR_2^1, S^1)),
    \end{multline*}
    otherwise, $ \indicator({\IncR_1^1 \in \pubaction{j}{1}}) = 0$.
    Therefore, the total revenue of publisher $\prover_j$ and validator $V^1$ when following the Prrr protocol is:
    \begin{multline}\label{eq:revenue_colluder_follow_protocol}
        R_{\textit{collude}} (\textbf{\report}_{Pj}, \mathbf{0},\preinc, S^k, \puballrev{j}(\mathcal{G}^{2})) =\\
        \begin{cases}
            \max\limits_{\report \in \bigcup_{i} \textbf{\report}_{Pi}} \rewardr(\report, S^1), \\
            \;\;\;\;\;\;\; \text{if }\argmax\limits_{\report \in \bigcup_{i} \textbf{\report}_{Pi}} \rewardr(\report, S^1) \in \textbf{\report}_{Pj}; \\\\
            \maxtwo{\report \in \bigcup_{i} \textbf{\report}_{Pi}} \rewardr(\report, S^1), \hfill \;\;\;\; \text{otherwise.}
        \end{cases}
    \end{multline}

    (3) Revenue with Deviation:

    We now analyze the revenue of the colluding parties when they deviate from the Prrr protocol.
    We separate into two cases based on whether validator $V^1$ includes any report.

    Case (i): Validator $V^1$ includes no report, i.e., ${\IncludeF^1 = ()}$.

    In this case, the total revenue (Equation~\ref{eq:collude_revenue_structure}) of publisher~$\prover_j$ and validator $V^1$ is $\puballrev{j}(\mathcal{G}^{2})$.
    Since there is no colluding validator in subgame $\mathcal{G}^{2}$, all participants following the Prrr protocol in subgame $\mathcal{G}^{2}$ constitute an SPNE(Theorem~\ref{theorem:ic}).
    The deviation of publisher~$\prover_j$ does not increase her expected revenue in subgame $\mathcal{G}^{2}$.
    Therefore, the revenue of publisher~$\prover_j$ in subgame $\mathcal{G}^{2}$ is no higher than her revenue by following the Prrr protocol.
    Recall that by following the Prrr protocol in subgame $\mathcal{G}^{2}$, the revenue of publisher is proportional to the ratio of her report count to the total report count, multiplied by the total expected revenue (Observation~\ref{observation:publisher:allreports_nobribery}).
    Thus, we have:
    \begin{multline*}
        \puballrev{j}(\mathcal{G}^{2}) \leq \frac{|\textbf{\report}_{Pj}|}{\sum_{1 \leq i \leq n} |\textbf{\report}_{Pi}|} \cdot  \TotalPubR\left(\sum_{1 \leq i \leq n} |\textbf{\report}_{Pi}|\right) \,\,.
    \end{multline*}

    Due to Skipping Resistance (Property~\ref{property:bribery_resistance}), the total reward for all publishers ${\TotalPubR\left(\sum_{1 \leq i \leq n} |\textbf{\report}_{Pi}|\right)}$ is upper bounded by the minimum value of the reports.
    Consequently, the revenue of publisher $\prover_j$ in subgame $\mathcal{G}^{2}$ cannot exceed the total revenue of the colluding parties when they follow the Prrr protocol (Equation~\ref{eq:revenue_colluder_follow_protocol}).
    Therefore, in this case, the combined revenue of publisher $\prover_j$ and validator~$V^1$ with deviation is no higher than their revenue when adhering to the Prrr protocol.

    Case (ii): Validator $V^1$ includes at least one report.

    We analyze two sub-cases based on whether validator~$V^1$ includes a report from publisher $\prover_j$ as the first report.

    Sub-case (ii.a): Validator $V^1$ includes a report from publisher $\prover_j$ as the first report, i.e., $\IncR_1^1 \in \pubaction{j}{1}$.

    In this sub-case, the total revenue (Equation~\ref{eq:collude_revenue_structure}) of publisher $\prover_j$ and validator $V^1$ is no higher than the private value of $\prover_j$'s report included as the first report:
    \begin{align*}
        &R_{\textit{collude}} (\pubaction{j}{1}, \brbaction{j}{1},\IncludeF^1, S^1, \puballrev{j}(\mathcal{G}^{2}))\\
        =&(\rewardr(\IncR_1^1, S^1)- \rewardr(\IncR_2^1, S^1)) + \\
        &\indicator({|\IncludeF^1|=2}) \cdot \rewardr(\IncR_2^1, S^1)\\
        \leq& \rewardr(\IncR_1^1, S^1) \,\,.
    \end{align*}
    
    Hence, the total revenue of the colluders is no higher than the maximum private value of $\prover_j$'s report with random string $S^1$:
    \begin{multline*}
        R_{\textit{collude}} (\pubaction{j}{1}, \brbaction{j}{1},\IncludeF^1, S^1, \puballrev{j}(\mathcal{G}^{2})) \leq \\
        \max_{\report \in \textbf{\report}_{Pj}} \rewardr(\report, S^1) \,\,.
    \end{multline*}

    We now turn back to the revenue of colluders when following the Prrr protocol (Equation~\ref{eq:revenue_colluder_follow_protocol}).
    If publisher $\prover_j$ holds the report with the highest value among all publishers, then the total revenue of colluders by following the Prrr protocol is the maximum private value of $\prover_j$'s report with random string $S^1$.
    Otherwise, the total revenue of colluders by following the Prrr protocol is the second-highest private value among all publishers, which is no less than the maximum private value of $\prover_j$'s report.
    Therefore, in this sub-case, the combined revenue of publisher $\prover_j$ and validator~$V^1$ with deviation is no higher than their revenue when adhering to the Prrr protocol.

    Sub-case (ii.b): Validator~$V^1$ does not include any report from publisher $\prover_j$ as the first report, i.e., $\IncR_1^1 \notin \pubaction{j}{1}$.

    In this sub-case, the total revenue (Equation~\ref{eq:collude_revenue_structure}) of publisher $\prover_j$ and validator~$V^1$ is no higher than the private value of the second report included by validator~$V^1$:
    \begin{align*}
        &R_{\textit{collude}} (\pubaction{j}{1}, \brbaction{j}{1},\IncludeF^1, S^1, \puballrev{j}(\mathcal{G}^{2}))  \\
        =&\indicator({|\IncludeF^1|=2}) \cdot \rewardr(\IncR_2^1, S^1) \\
        \leq& \rewardr(\IncR_2^1, S^1) \,\,.
    \end{align*}

    When there are multiple reports with the same highest value, the revenue of colluders with deviation is no higher than that highest value.
    At the same time, the revenue of colluders by following the Prrr protocol (Equation~\ref{eq:revenue_colluder_follow_protocol}) is exactly that highest value since the highest and second-highest values are equal.

    When there is only one report with the highest value, if the validator~$V^1$ includes that report as the second report, then the random value of the second report is higher than the first.
    Recall that in Section~\ref{sec:game:progress}, we introduce a dummy report with the minimum random value as the second report in the above case.
    Consequently, the revenue of the colluding parties with deviation is at most the second-highest private value among all publishers. 
    However, the revenue of the colluding parties when following the Prrr protocol~(Equation~\ref{eq:revenue_colluder_follow_protocol}) is at least this second-highest value. 
    Thus, in this sub-case, the combined revenue of publisher~$\prover_j$ and validator~$V^1$ with deviation is no higher than their revenue when adhering to the Prrr protocol.

    (4) Conclusion:
    In all, we have shown that in all possible deviation cases, the combined revenue of publisher $\prover_j$ and validator~$V^1$ with deviation is no higher than their revenue when adhering to the Prrr protocol.
    Thus, collusion between any publishers and validator~$V^1$ in the first step does not increase their total revenue.
\end{maybeProof}

\subsubsection{Equilibrium Stability}\label{sec:ic:robustness:stability}

Indeed, we saw (\S\ref{sec:ic:eq}) that Prrr is a (non-strict) SPNE.
We prove that validator deviations do strictly reduce their revenue if Prrr employs the logarithmic random-value function (\S~\ref{appendix:validator_deviation}). 
We now show that if a publisher changes (either increases or decreases) the revenue of the validator or of other publishers, her own revenue strictly decreases.

Publishers may also have multiple best responses when all participants follow the Prrr protocol.
We prove that no publisher can deviate in a way that changes the revenue of other publishers unless the deviation also reduces their own revenue.

\begin{lemma}\label{lemma:equilibrium_publisher_stability}
    In the game $\mathcal{G}(\rewardr)$, assuming the random-value function $\rewardr$ satisfies Reward Monotonicity (Property~\ref{property:monotonicity}) and Skipping Resistance (Property~\ref{property:bribery_resistance}).
    Suppose all publishers follow the Prrr protocol except for a publisher $\prover_j$ who deviates by taking action $(\pubaction{j}{k}, \brbaction{j}{k})$ in step $k$.
    Assume that if following the Prrr protocol is a best response for a validator $V^k$ then she follows the Prrr protocol.
    If the revenue of any validator or any other publisher changes because publisher~$\prover_j$ deviates, then the expected revenue of publisher $\prover_j$ strictly decreases.
\end{lemma}

We first show that if the best response of the validator to the deviation is to include no report, then the expected revenue of the deviating publisher strictly decreases.
Next, we assume that the best response of the validator is to include some report.
We prove that if the publisher~$\prover_j$ does not publish all her reports, then her expected revenue strictly decreases (Lemma~\ref{lemma:publisher:deviation_revenue_compreto_nobr}).
Finally, in the case where publisher~$\prover_j$ publishes all her reports but the revenue of any other participants changes, we prove that the expected revenue of publisher~$\prover_j$ strictly decreases.
\begin{showWhenSubmit}The proof is in the full version of this report~\cite{anonymous2025prrr}.\end{showWhenSubmit}

\begin{maybeProof}
    Since if all participants follow the Prrr protocol the game ends at the first step, we primarily focus on the first step.
    We denote by $(\pubaction{j}{1},\brbaction{j}{1})$ the deviating action of publisher $\prover_j$ in step~$1$.
    We analyze two cases based on whether validator $V^1$'s best response is to include any report.

    Before diving into the cases, we first build up some common notation and results.

    Recall that $\puballrev{j}(\mathcal{G}^{2})$ is the expected revenue for publisher $\prover_j$ in the subgame starting from step 2.
    Since following Prrr constitutes an SPNE in all subgames (Theorem~\ref{theorem:ic}), the deviation of publisher~$\prover_j$ does not increase her expected revenue in subgame $\mathcal{G}^{2}$.
    Therefore, the revenue of publisher~$\prover_j$ in subgame $\mathcal{G}^{2}$ is no higher than the total expected revenue of publishers if they follow the Prrr protocol, that is: 
    \begin{equation}\label{eq:publisher:deviation:subgame_revenue_upperbound}
        \puballrev{j}(\mathcal{G}^{2}) \leq \TotalPubR\left(\sum_{1 \leq i \leq n} |\textbf{\report}_{Pi}|\right)
    \end{equation}

    Now, we analyze the revenue of publisher $\prover_j$ with deviation if the best response of validator $V^1$ is to include some reports, that is, $\valbr^1(\cdot) \neq ()$.
    Under this condition, given the random string $S^1$, the revenue of publisher $\prover_j$ in the game with deviation (Equation~\ref{eq:publisher:utility}) is just her revenue in step $1$:
    \begin{multline*}
        \pubsinglesteprev{j}{1\sim m}(\pubaction{j}{1}, \brbaction{j}{1},\IncludeF^1 \neq (), S^1, \puballrev{j}(\mathcal{G}^{2})) = \\
        \pubsinglesteprev{j}{1}(\pubaction{j}{1}, \brbaction{j}{1},\IncludeF^1, S^1) \,\,.
    \end{multline*}

    By Lemma~\ref{lemma:publisher:deviation_revenue_compreto_nobr}, for any random string $S^1$, the revenue of publisher $\prover_j$ in step $1$ with deviation is at most her revenue when she takes action $(\textbf{\report}_{Pj}, \mathbf{0})$ and validator $V^1$ follows the Prrr protocol.
    Since her total revenue in the game is just her revenue in step $1$, it follows that with any deviating action $(\pubaction{j}{1}, \brbaction{j}{1})$, her revenue is no higher than if she takes action $(\pubaction{j}{1}, \mathbf{0})$ and the validator follows the Prrr protocol:
    \begin{multline}\label{eq:publisher:deviation:revenue_upperbound}
        \pubsinglesteprev{j}{1\sim m}(\pubaction{j}{1}, \brbaction{j}{1},\IncludeF^1 \neq (), S^1, \puballrev{j}(\mathcal{G}^{2})) \leq \\
        \pubsinglesteprev{j}{1\sim m}(\pubaction{j}{1}, \mathbf{0},\preinc, S^1, \puballrev{j}(\mathcal{G}^{2})) \,\,.
    \end{multline}

    Now we consider the two cases separately.

    Case (1): There is a random string $S^1$ where validator~$V^1$'s best response to the deviation is to include no report.

    In this case, we prove that the expected revenue of publisher~$\prover_j$ with deviation is strictly less than her expected revenue by following the Prrr protocol.

    We compare the revenue of publisher $\prover_j$ with deviation to the revenue she would obtain by the same set of reports without bribes, i.e., $(\pubaction{j}{1}, \mathbf{0})$.
    It suffices to show that her revenue with deviation if the validator follows any best response is strictly less than her revenue with action $(\pubaction{j}{1}, \mathbf{0})$ if the validator follows the Prrr protocol; in this case, the deviation cannot be a best response.
    Note that with $\prover_j$'s action $(\pubaction{j}{1}, \mathbf{0})$, following the Prrr protocol is a best response for validator $V^1$.

    If the best response of the validator is to include no report, then the revenue of the validator only comes from bribes (Equation~\ref{eq:validator:utility}).
    Since following the Prrr protocol has at least $\minr$ revenue for the validator, the total bribe offered by all publishers must be at least $\minr$.
    Thus, the publisher~$\prover_j$ must pay at least $\minr$ bribe.
    Therefore, her revenue in the game $\mathcal{G}$ is most $\puballrev{j}(\mathcal{G}^{2}) - \minr$.

    Combining Skipping Resistance (Property~\ref{property:bribery_resistance}) and Equation~\ref{eq:publisher:deviation:subgame_revenue_upperbound}, we have $\puballrev{j}(\mathcal{G}^{2}) - \minr < 0$.
    Therefore,~$\prover_j$'s revenue in this case is strictly less than $0$, which is also strictly less than her revenue when she takes action $(\pubaction{j}{1}, \mathbf{0})$ and the validator follows the Prrr protocol:
    \begin{multline*}
        \pubsinglesteprev{j}{1\sim m}(\pubaction{j}{1}, \brbaction{j}{1},\IncludeF^1 = (), S^1, \puballrev{j}(\mathcal{G}^{2})) < \\ \pubsinglesteprev{j}{1\sim m}(\pubaction{j}{1}, \mathbf{0},\preinc, S^1, \puballrev{j}(\mathcal{G}^{2}))
    \end{multline*}

    Recall that if the best response of the validator is to include some reports, the revenue of publisher $\prover_j$ with deviation is at most her revenue when she takes action $(\pubaction{j}{1}, \mathbf{0})$ and the validator follows the Prrr protocol (Equation~\ref{eq:publisher:deviation:revenue_upperbound}).
    Therefore, if there exists any random string~$S^1$ for which the validator's best response is to include no report, the expected revenue of publisher~$\prover_j$ with deviation is strictly less than her expected revenue when taking action $(\pubaction{j}{1}, \mathbf{0})$ and the validator follows the Prrr protocol.
    Thus, such a deviation cannot be a best response for publisher~$\prover_j$.

    Case (2): For all random strings $S^1$, validator $V^1$'s best response to the deviation is to include some reports.

    We separate to two subcases, based on whether the publisher $\prover_j$ publishes all of her reports.

    Subcase (i), Publisher $\prover_j$ does not publish all of her reports, i.e., $\pubaction{j}{1} \neq \textbf{\report}_{Pj}$:

    In this subcase, we prove that the expected revenue of publisher $\prover_j$ with deviation is strictly less than her expected revenue by following the Prrr protocol.

    Recall that in this case including an empty set of reports is not the best response of validator $V^1$.
    Therefore, the revenue of publisher $\prover_j$ with deviation is at most her revenue when she takes action $(\pubaction{j}{1}, \mathbf{0})$ and the validator follows the Prrr protocol (Equation~\ref{eq:publisher:deviation:revenue_upperbound}).
    Therefore the expected revenue of publisher $\prover_j$ with deviation if the validator follows any best response is at most her expected revenue when taking action $(\pubaction{j}{1}, \mathbf{0})$ and the validator follows the Prrr protocol.
    Combined with Observation~\ref{observation:revenue_no_bribery}, we have:
    \begin{align*}
        &\mathbb{E}_{S^1}\left[\pubsinglesteprev{j}{1\sim m}(\pubaction{j}{1}, \brbaction{j}{1},\IncludeF^1 \neq (), S^1, \puballrev{j}(\mathcal{G}^{2}))\right] \\
        \leq &\mathbb{E}_{S^1}\left[\pubsinglesteprev{j}{1\sim m}(\pubaction{j}{1}, \mathbf{0},\preinc, S^1, \puballrev{j}(\mathcal{G}^{2}))\right]\\
        =& \frac{|\pubaction{j}{1}|\cdot  \TotalPubR\left(|\pubaction{j}{1}| + \sum_{i\neq j} |\textbf{\report}_{Pi}|\right)}{|\pubaction{j}{1}| + \sum\limits_{i \neq j} |\textbf{\report}_{Pi}|}  \,\,.
    \end{align*}

    While by following the Prrr protocol, the expected revenue of publisher $\prover_j$ is:
    \begin{multline*}
        \mathbb{E}_{S^1}\left[\pubsinglesteprev{j}{1\sim m}(\textbf{\report}_{Pj}, \mathbf{0},\preinc, S^1, \puballrev{j}(\mathcal{G}^{2}))\right] = \\
        \frac{|\textbf{\report}_{Pj}|}{\sum_{1 \leq i \leq n} |\textbf{\report}_{Pi}|} \cdot  \TotalPubR\left(\sum_{1 \leq i \leq n} |\textbf{\report}_{Pi}|\right) \,\,.
    \end{multline*}
    Since $|\pubaction{j}{1}| < |\textbf{\report}_{Pj}|$, due to Reward Monotonicity (Property~\ref{property:monotonicity}), we have:
    \begin{multline*}
         \TotalPubR\left(|\pubaction{j}{1}| + \sum_{i\neq j} |\textbf{\report}_{Pi}|\right) \leq \\ \TotalPubR\left(\sum_{1 \leq i \leq n} |\textbf{\report}_{Pi}|\right) \,\,.
    \end{multline*}
    We also trivially have:
    \begin{equation*}
        \frac{|\pubaction{j}{1}|}{|\pubaction{j}{1}| + \sum\limits_{i \neq j} |\textbf{\report}_{Pi}|} < \frac{|\textbf{\report}_{Pj}|}{\sum_{1 \leq i \leq n} |\textbf{\report}_{Pi}|} \,\,. \\
    \end{equation*}
    Therefore, the expected revenue of publisher $\prover_j$ with deviation is strictly less than her expected revenue by following the Prrr protocol.

    Subcase (ii), Publisher $\prover_j$ publishes all of her reports, i.e., ${\pubaction{j}{1} =\textbf{\report}_{Pj}}$:

    In previous case and subcase, we have only shown that the expected revenue of publisher $\prover_j$ with deviation is strictly less than her expected revenue by following the Prrr Protocol.
    In this subcase, we study whether the deviation of publisher $\prover_j$ affects the revenue of other participants.

    First, the revenue of validators after step $1$ remains unaffected since validator~$V^1$ includes some reports in step~$1$~(condition of Case (2)).
    The revenue of validators after step $1$ consistently remains zero, regardless of whether a deviation occurs.

   Second, if the deviating publisher~$\prover_j$ could increase the revenue of the first validator~$V^1$ without reducing her own revenue, it would imply that their collusion increases their total revenue. 
    This contradicts Lemma~\ref{lemma:collusion:pubval_no_increase}.
    On the other hand, since~$\prover_j$ still publishes all reports in the deviation (as per the condition of this subcase), the validator~$V^1$, by following the Prrr protocol, would receive the bribe from~$\prover_j$ plus the second-highest value. 
    This is no less than the revenue obtained by following the Prrr protocol when~$\prover_j$ does not deviate.
    Thus, $\prover_j$ cannot reduce the revenue of validator~$V^1$.
    In conclusion, $\prover_j$ cannot alter validator~$V^1$'s revenue without reducing her own revenue.

    Therefore, the remaining task is to prove that the revenue of other publishers would not be affected without reducing~$\prover_j$'s revenue.

    To change the revenue of another publisher $\prover_{j\prime}$, the deviation of publisher $\prover_j$ must change the revenue of $\prover_{j\prime}$ with at least one random string $S^1$.
    We prove that, in this case, the revenue of publisher $\prover_j$ with deviation is strictly less than her revenue when following the Prrr protocol.

    To make following the Prrr protocol not a best response for validator $V^1$ with random string $S^1$, publisher $\prover_j$ needs to at least provide a positive bribe.
    Therefore, if the best response of validator $V^1$ to deviation does not include the report from publisher $\prover_{j}$ as the first report, $\prover_{j}$'s revenue with deviation is strictly less than zero.
    Thus, if~$\prover_{j}$'s revenue does not decrease, the best response of validator $V^1$ to deviation must include the report from publisher $\prover_{j}$ as the first report.

    If validator $V^1$ does not include $\prover_{j\prime}$'s report as the first report when all participants follow the Prrr protocol, then~$\prover_{j\prime}$'s revenue is zero, which remains unchanged with any deviation. 
    Therefore, for $\prover_{j\prime}$'s revenue to change due to a deviation, validator $V^1$ must include $\prover_{j\prime}$'s report as the first report when all participants follow the Prrr protocol. 
    This implies that $\prover_{j\prime}$'s report has the unique highest value among all reports; otherwise, if the highest value is not unique, $\prover_{j\prime}$'s revenue is zero according to the Prrr protocol, contradicting the assumption that her revenue changes with deviation.

    Therefore, we assume that without deviation, validator~$V^1$ includes the report from publisher $\prover_{j\prime}$ as the first report.

    Recall that $\max^{(2)}_{\report \in \bigcup_{i} \textbf{\report}_{Pi}} \rewardr(\report, S^1)$ is the second-highest value among all reports with random string $S^1$.
    The revenue of the validator when all participants follow the Prrr protocol is exactly the value of this second-highest report:
    \begin{equation*}
        R_V^1(\preinc,\textbf{0}, S^1) = \max^{(2)}_{\report \in \bigcup_{i} \textbf{\report}_{Pi}} \rewardr(\report, S^1) \,\,.
    \end{equation*}

    Denote by $r_1$ and $r_2$ the private values of the reports validator $V^1$ includes as the first and second reports to deviation, respectively.
    The revenue of validator $V^1$ with deviation is $r_2$ plus the bribe from publisher $\prover_j$:
    \begin{equation*}
        R_V^1(\valbr^1(\cdot),\brbaction{j}{1}, S^1) = r_2 + \brbaction{j}{1}(\valbr^1(\cdot), S^1) \,\,.
    \end{equation*}
    This revenue must be strictly higher than the revenue when all participants follow the Prrr protocol. Otherwise, following the Prrr protocol would still be a best response for validator $V^1$ with deviation.
    Therefore, we have the lower bound on the bribe offered by publisher $\prover_j$:
    \begin{equation}\label{eq:publisher:deviation:bribe_lowerbound}
        \brbaction{j}{1}(\valbr^1(\cdot), S^1) > \max^{(2)}_{\report \in \bigcup_{i} \textbf{\report}_{Pi}} \rewardr(\report, S^1) - r_2\,\,.
    \end{equation}

    The revenue of publisher $\prover_j$ with deviation is:
    \begin{align*}
        &\pubsinglesteprev{j}{1}(\textbf{\report}_{Pj}, \brbaction{j}{1},\IncludeF^1, S^1) \\
        =& (r_1 - r_2) - \brbaction{j}{1}(\IncludeF^1) \\
        \stackrel{Eq. \ref{eq:publisher:deviation:bribe_lowerbound}}{<} & (r_1 - r_2) - \left(\max^{(2)}_{\report \in \bigcup_{i} \textbf{\report}_{Pi}} \rewardr(\report, S^1) - r_2\right) \\
        =& r_1 - \max^{(2)}_{\report \in \bigcup_{i} \textbf{\report}_{Pi}} \rewardr(\report, S^1) \,\,.
    \end{align*}

    Since $r_1$ is a report from publisher $\prover_j$ but not the report with the highest value, we have $$r_1 \leq \max^{(2)}_{\report \in \bigcup_{i} \textbf{\report}_{Pi}} \rewardr(\report, S^1).$$
    Therefore, the revenue of publisher $\prover_j$ with deviation with $S^1$ is strictly negative.
    This implies that any such deviation strictly reduces her revenue compared to following the Prrr protocol.

    From Equation~\ref{eq:publisher:deviation:revenue_upperbound}, with any random string $S^1$, publisher~$\prover_j$'s revenue with any deviating action $(\textbf{\report}_{Pj}, \brbaction{j}{1})$ if the validator follows any best response is not higher than her revenue when she follows the Prrr protocol.
    Therefore, if there exists any random string $S^1$ where the revenue of publisher~$\prover_j$ with deviation is strictly less than her revenue when following the Prrr protocol, then the deviation is not the best response for the publisher.
    
    \textbf{Conclusion}
    In all, considering all cases, we have shown that if the deviation of publisher~$\prover_j$ changes the revenue of any other publisher, then the expected revenue of publisher~$\prover_j$ strictly decreases.
\end{maybeProof}

\section{Conclusion}\label{sec:conclusion}
We present Prrr, a novel protocol for the prevalent blockchain reporting problem that addresses the security-performance trade-off.
Prrr utilizes a mechanism-design concept called EASA, which is the first game-theoretic concept to deliberately form participant asymmetry.
We prove that all participants following the Prrr protocol constitutes a subgame perfect Nash equilibrium (SPNE) and this SPNE is robust against collusion and Sybil attacks.
Prrr is directly applicable to the numerous smart contracts that rely on timely reports.

\section*{Acknowledgements}

This work was supported in part at NSFC/RGC Joint Research Scheme under Grant 62261160391 and Grant N\_PolyU529/22 and at the Technion by an Aly Kaufman Fellowship, IC3 and Avalanche Foundation.

\bibliography{ref}

\appendices

\section{Ethereum Reports Data}\label{appendix:ethereum_data}

We collect approximately 30 days of Ethereum data from 15/01/2026 to 13/02/2026 (block \num{24236568} to block \num{24449610}).
During this period, we observe an average of approximately \num{1530} state update and batch update reports per day across 8 rollups: about 1250 Arbitrum~\cite{Arbitrum}, 24 Base~\cite{base_node}, 24 Optimism~\cite{optimism}, 20 Linea~\cite{linea_docs}, 85 Polygon~\cite{polygonzkEVMDocs}, 33 Scroll~\cite{scroll}, 71 Starknet~\cite{starknet2023}, and 22 ZkSync~\cite{zksync}.
Additionally, we observe around \num{260} liquidation calls per day across 5 lending protocols: 35 Aave~v2~\cite{aave_v2_docs}, 190 Aave~v3~\cite{aave_v3_overview}, 8 Compound~v3~\cite{compound_v3_docs}, 20 Morpho~\cite{morpho_docs}, and 11 Spark~\cite{spark_docs}.

On average, the gas prices (including both base fee and tips) for these reports are approximately $13.8$ times of the median gas price in their respective blocks. 
Note that this calculation excludes any bribes beyond gas fees.

\section{Prrr Deployment}\label{appendix:practical_conversions}

We illustrate Prrr's applicability with two examples.

The first example is PoW reporting for zk-rollups~\cite{wang2025p}. 
Currently, these systems rely on centralized publishers to periodically publish zk-proofs for state updates.
To decentralize this process, zk-proofs can be treated as PoW reports (cf. Aleo~\cite{aleo2025posvpow}'s consensus that uses zk-proofs as PoW).
In each epoch, the smart contract specifies a report-generation window during which publishers generate zk-proofs (or leverages decentralized prover markets~\cite{wang2025p}), followed by a report-publication window.
Prrr is then used to reward publishers in the layer-1 contract.

Another example is PoS reporting for watchtowers~\cite{xu2023silentower,avarikioti2020cerberus,liu2020fail,khabbazian2019outpost}, which monitor the behavior of off-chain entities and report any misbehavior to the blockchain.
In this scenario, reports typically involve only a few hash operations and signature verifications, making misbehavior report generation computationally efficient.
Consequently, any publisher can generate a report within a very short time, and multiple reports can leverage the same misbehavior evidence.
Therefore, a report-generation window is unnecessary.
When misbehavior arises off-chain, a reporting epoch begins and immediately enters the report-publication window.
Publishers publish misbehavior reports, each accompanied by a deposit.
Prrr then allocates rewards to publishers.

\section{Practical Considerations of Prrr}\label{appendix:practical}

First, some blockchain protocols, such as Ethereum~\cite{buterin2013ethereum}, require a minimum fee for each transaction. 
This minimum fee is not awarded to validators but is instead \emph{burned}, i.e., permanently removed from circulation. 
The Prrr protocol can simply add an amount equal to the minimum fee to the reward paid to the publishers of the (up to) two included reports.
Thus, the minimum fee would not affect our analysis. 
Since the protocol includes at most two reports per block, the additional cost is nominal.

Second, most blockchain protocols process transactions within a block sequentially. 
When executing a transaction, the system cannot access information about subsequent transactions in the same block. 
As a result, when processing a transaction containing a report, the smart contract cannot immediately determine whether it is the last report in the block and allocate the reward accordingly. 
To address this, the smart contract processes the first report as usual and records any additional reports for later evaluation. 
The final reward calculation can then be performed after all relevant information is available.

Third, in practical systems, publishers can specify a validity duration window for each transaction, eliminating the need to publish transactions at every step. 
If a publisher publishes a transaction with a duration window spanning multiple steps, we can simply treat it as if the transaction is published for each step within that window. 
Then, it does not affect the analysis of our protocol.
In our protocol, publishers can publish all report transactions by specifying a transaction window that encompasses all steps within the report-publication window, achieving the same outcome.

\section{Analysis of Random-Value Functions}\label{appendix:value_function}

We analyze the logarithmic (\S\ref{appendix:value_function:log}) and polarized (\S\ref{appendix:value_function:polarized}) random-value functions separately.
For each case, we first discuss parameter selection to ensure Reward Monotonicity (Property~\ref{property:monotonicity}) and Skipping Resistance (Property~\ref{property:bribery_resistance}).
Then, we analyze the expected cost incurred by the smart contract when using the random-value function.

\subsection{The Logarithmic Random-Value Function}\label{appendix:value_function:log}

Recall that $\minr$ is the minimum of the random-value function, $H(\cdot)$ is the random oracle, $\lambda > 0$ is a parameter that controls the distribution of rewards.
Given a report $\report$ and a random string $S$, recall that the logarithmic random-value function is (\S\ref{sec:mechanism:valuefunctions}):
\begin{equation*}
    \rewardr_{\textit{log}}(\report, S) = \minr - \frac{1}{\lambda}\ln(1 - H(\report || S)) \,\,.
\end{equation*}

Then the distribution of $\rewardr_{\textit{log}}(\report, S)-\minr$ follows an exponential distribution with parameter $\lambda$:
\begin{multline*}
    \Pr(\rewardr_{\textit{log}}(\report, S) - \minr \geq r ) 
    = \\ \Pr\left(-\frac{1}{\lambda}\ln(1- H(\report || S)) \geq r\right) = e^{-\lambda r}
\end{multline*}

For any $N$ reports $\{\report_i\}_{i=1}^N$ and any random string~$S$, the random variables $\rewardr_{\textit{log}}(\report_i,S)-\minr$ are i.i.d. exponential random variables with parameter $\lambda$.
By the memoryless property of the exponential distribution, the difference between the highest and second-highest values also follows an exponential distribution with parameter $\lambda$~\cite{david2004order}.
Therefore, the expected total reward for all publishers when there are $N$ reports is fixed to ${\TotalPubR(N) = \frac{1}{\lambda}}$, regardless of $N$.
Thus, Reward Monotonicity (Property~\ref{property:monotonicity}) is satisfied trivially.
To satisfy Skipping Resistance (Property~\ref{property:bribery_resistance}), we need $\TotalPubR(N_{\max}) = \frac{1}{\lambda} < \minr$.
Therefore, by setting~$\lambda > \frac{1}{\minr}$, we ensure that both properties are satisfied.

The expected reward for validators is the expected second-highest value among $N$ reports and the expected reward for publishers is the expected difference between the highest and second-highest value among $N$ reports.
Therefore, the total expected cost for the smart contract is the expected highest value among $N$ reports.

Denote by $H_N = \sum_{i=1}^N \frac{1}{i}$ the $N$-th harmonic number.
For $N$ i.i.d. random variables following the exponential distribution with parameter $\lambda$, the expected highest value among them is $\frac{H_N}{\lambda}$~\cite{david2004order}.
Therefore, for the logarithmic random-value function, the total expected cost for the smart contract when there are~$N$ reports is $\minr + \frac{H_N}{\lambda}$.
Since~$H_N$ is~$O(\log N)$, the total expected cost for the smart contract increases logarithmically with the number of reports.


\subsection{The Polarized Random-Value Function}\label{appendix:value_function:polarized}

Recall that $\minr$ and $\maxr$ are the two possible outcomes of the polarized random-value function, where~${\maxr > \minr}$,~$p$ is the probability of receiving $\maxr$, and~$H(\cdot)$ is the random oracle.
Given a report $\report$ and a random string~$S$, recall that the polarized random-value function is~(\S\ref{sec:mechanism:valuefunctions}):
\begin{equation*}
    \rewardr_{\textit{polarized}}(\report, S) = 
    \begin{cases} 
        \maxr & \text{if } H(\report || S) \leq p \\
        \minr & \text{otherwise}
    \end{cases}
\end{equation*}

Publishers have positive revenue only when there is exactly one report that has the high value $\maxr$.
When there are~$N$ reports in total, the probability is $N \times p \times (1-p)^{N-1}$.
Therefore, the expected total reward for all publishers is
\begin{equation}\label{eq:expected_reward}
    \TotalPubR(N) = N\times p(1-p)^{N-1}(\maxr - \minr)\,\,.
\end{equation}

For monotonicity, since $\frac{\TotalPubR(N)}{\TotalPubR(N-1)}  = \frac{N}{N-1} \times (1-p)$, $\TotalPubR(N)$ is monotonically increasing with $N$ for $N \leq \frac{1}{p}$.
Therefore, by setting $p \leq \frac{1}{N_{\max}}$, the monotonicity property is satisfied.

In addition, to satisfy Skipping Resistance (Property~\ref{property:bribery_resistance}), we need $\TotalPubR(N_{\max}) < \minr$ for all $N \leq N_{\max}$.
Due to monotonicity, it suffices to ensure $\TotalPubR(N_{\max}) < \minr$.
The condition becomes:
\begin{equation*}
    \maxr < (1 + \frac{1}{N_{\max} p (1-p)^{N_{\max}-1}}) \minr \,\,.
\end{equation*}

In all, by setting $p \leq \frac{1}{N_{\max}}$ and ${\maxr < (1 + \frac{1}{N_{\max} p (1-p)^{N_{\max}-1}}) \minr}$, both properties are satisfied for the polarized random-value function.

Recall that the expected total cost for the smart contract is the expected highest value among $N$ reports (\S\ref{appendix:value_function:log}).
For the polarized random-value function, the maximal expected cost to the smart contract is $\maxr$.
The expected cost to the smart contract when there are $N$ reports is the probability that at least one report has the high value $\maxr$ times $\maxr$ plus the probability that no report has the high value $\maxr$ times $\minr$:
\begin{align*}
    &E_{S}[\max(\max_{1\leq i \leq N} \rewardr_{\textit{polarized}}(\report_i, S), \minr)] 
    \\ =& (1 - (1-p)^N) \times \maxr + (1-p)^N \times \minr 
    \\ =& \minr + (1 - (1-p)^N) \times (\maxr - \minr)
\end{align*}

\section{Validator Deviations}\label{appendix:validator_deviation}
We analyze validator deviations when all publishers follow the Prrr protocol.
When multiple reports share the highest value, the validator has several best responses: she may place any two of these highest-value reports as the first and second reports. 
All such choices are consistent with the Prrr protocol and yield identical publisher revenue, since the reward, which is the difference between the highest and second-highest values, is zero.
If the highest value is unique but several reports share the second-highest value, the validator may select any two of these second-highest reports as the first and second included reports. 
This does not affect the validator's revenue, but it strictly reduces the publisher's revenue for the highest-value report to zero.
In all other cases, the validator's best response is unique: she can only obtain the second-highest value reward by following the Prrr protocol (Equation~\ref{eq:validator:utility}), and any deviation would strictly decrease her revenue.
\begin{observation}\label{observation:validator_strict_deviation}
    In the game $\mathcal{G}$, suppose all publishers follow the Prrr protocol.
    For a validator $V^k$, if the highest value among all published reports is unique and the second-highest value is also unique, then any deviation from the Prrr protocol strictly decreases the validator's revenue.
\end{observation}

For the logarithmic value function, the probability that two reports have exactly the same value is zero since it is a continuous distribution.
Therefore, we have 
\begin{corollary}
    In the game $\mathcal{G}(\rewardr_{\textit{log}})$ where $\rewardr_{\textit{log}}$ is the logarithmic random-value function, suppose all publishers follow the Prrr protocol.
    Then, with probability $1$, any deviation from the Prrr protocol strictly decreases the validator's revenue.
\end{corollary}

For the polarized value function, since there are only two possible reward values, if there are at least three reports published, the validator always has multiple best responses when all publishers follow the Prrr protocol.

\end{document}